\documentclass{JHEP3}
\usepackage{multicol,bbm}
\usepackage{graphicx,psfrag}
\usepackage{dcolumn}
\usepackage{bm}
\usepackage{amsmath,amssymb}

\newcommand{\eq}{\begin{equation}}
\newcommand{\eqe}{\end{equation}}

\newcommand{\eqa}{\begin{eqnarray}}
\newcommand{\eqae}{\end{eqnarray}}

\JHEPspecialurl{http://jhep.sissa.it/JOURNAL/JHEP3.tar.gz}

\title{Cosmology of the selfaccelerating third order Galileon}

\author{David F.~Mota\\
Institute of Theoretical Astrophysics, University of Oslo,
0315 Oslo, Norway}
\author{Marit Sandstad\\
Institute of Theoretical Astrophysics, University of Oslo,
0315 Oslo, Norway}
\author{Tom Zlosnik\\
Perimeter Institute for Theoretical Physics, 31 Caroline Street North, Waterloo, Ontario N2L 2Y5, Canada}

\received{\today} 
\accepted{} 

\abstract{In this paper we start from the original formulation of the galileon model with the original choice for couplings to gravity. Within this framework we find that there is still a subset of possible Lagrangians that give selfaccelerating solutions with stable spherically symmetric solutions. This is a certain constrained subset of the third order galileon which has not been explored before. We develop and explore the background cosmological evolution of this model drawing intuition from other even more restricted galileon models. The numerical results confirm the presence of selfacceleration, but also reveals a possible instability with respect to galileon perturbations.}

\preprint{}

\begin{document}

  \section{Introduction}
  
  Supernova data \cite{Riess_et_al1998, Perlmutter_et_al1999} in conjunction with data from the CMB \cite{Komatsu_et_al2010} suggests that the Universe is expanding at an accelerated rate. Though alternatives to accelerated expansion have been proposed in terms of inhomogeneous models (see for instance \cite{Zehavi_et_al1998}), it is now most commonly asked in cosmology not whether, but how or why the Universe's expansion is accelerated.

  The simplest way to get accelerated expansion is by adding a cosmological constant. This solution is now part of the standard cosmological model and it fits well with modern precision data of the cosmic microwave background, supernova data and baryon acoustic oscillations \cite{Komatsu_et_al2010}.

  However simple and elegant the cosmological constant solution may seem, questions still remain. If this really is the right explanation for the cosmic acceleration, what is the physical origin of the constant? Why has it got the value that it has? A possible physical source is vacuum energy, but attempts to predict its value from known microscopic physics notoriously yield results more than a hundred orders of magnitude larger than the observed values \cite{Weinberg1989}. Though there are proposed resolutions to this, appealing both to our lack of knowledge of fundamental physics and to anthropic arguments \cite{Weinberg1989}, one might alternatively look for a dynamical solution important on cosmological scales.

  The alternatives to the cosmological constant come in roughly two categories. The first proposes some sort of generalised dark energy \cite{Copeland_et_al2006}. The other possibility is that of some kind of modification of gravity on large scales \cite{ruth}. In most cases models of dark energy and modified gravity are dual in the sense that a conformal transformation can take a theory of one type into a theory of the other kind. The main difference lies in what the formulation is like in the frame where one assumes the background metric. If the theory in this case is minimally coupled to gravity, the theory is considered a dark energy model, whereas if it is nonminimally coupled to gravity in this frame it is considered a modified gravity model.
  
  The latter alternative consists of such alternatives as additions of generalised metric functions, $f(R)$-gravities \cite{fr} and  the Gauss-Bonnet-gravities \cite{Carroll_et_al2005,Koivisto:2006}, scalar-vector-tensor-theories \cite{Bekenstein_2004,Bour}, additions due to extra dimensions \cite{Deffayet_et_al2002} such as the DGP model \cite{Dvali_et_al2000}, \cite{Lue2005} and its generalisations, degravitation \cite{Dvali_et_al2007} and cascading gravities \cite{Agarwal_et_al2010}, the Brans-Dicke theories \cite{Farajollahi_et_al2010,deFelice_et_Tsujikawa2010c,Mota:2006} and many more.

  Recently the Galileon model \cite{Nicolis_et_al2008} is a modified gravity that has attracted a good deal of attention, though it has also been formulated as a pure dark energy candidate. Motivated by the DGP and its Vainshtein mechanism, the authors of \cite{Nicolis_et_al2008} constructed the Galileon model, a model of a scalar field $\pi$ that has a symmetry of the equations of motion under the galilean shift:
  \begin{equation}
    \pi \to \pi + c + b_\mu x^\mu
  \end{equation}
  and found all five possible Lagrangians that obey this symmetry in four dimensions. The five different terms are denoted by the order of the scalar field $\pi$ they contain, and there is one term unique up to total derivatives of each of the orders one to five, and none of higher than fifth order. They considered a theory built from linear combinations of these terms and found that it could yield selfaccelerating solutions in a cosmological background. By considering spherically symmetric solutions and their stability, bounds were put on the coefficients.

  After this promising start considerations of the cosmological evolutions of galileon theories or subsets of such theories were considered in \cite{Chow_et_Khoury}, which dealt with only the behaviour of the third order galileon term, which turned out to give a slight degravitation of the cosmological constant, but gave no viable selfaccelerating solutions, and for slightly different couplings to gravity than those given in \cite{Nicolis_et_al2008} in \cite{Gannouji_et_Sami2010, deFelice_et_Tsujikawa2010a, Ali_et_al2010} which all explored the cosmology of theories of up to fourth and in the case of \cite{deFelice_et_Tsujikawa2010a} fifth order galileon models. 

  Several other theoretical aspects of the galileon theory have also been considered, such as the null-energy-condition (NEC) violation \cite{Nicolis_et_al2009}, covariantisation \cite{Deffayet_Esposito_et_Vikman2009}, the Vainshtein mechanism and its effect on spherically symmetric solutions \cite{Burrage_et_Seery2010} and black hole accretion \cite{Babichev2010}.

  From theoretical considerations of the symmetries, many generalisations have been built. In \cite{Deffayet_Deser_et_Esposito2009} expressions for all the galileon invariant Lagrangians in arbitrary dimensions were derived. Generalisations to arbitrary p-form galileons were obtained in \cite{Deffayet_Deser_et_Esposito2010} and the multi-galileons, which are multiplets of scalar galileons, contained in this formalism has since been considered in \cite{Padilla_et_al2010a} and \cite{Padilla_et_al2010b} which mainly focused on the two- or bi-galileon case, in \cite{ Padilla_et_al2010c} which focused on theoretical aspects and the evasion of Derrick's theorem \cite{Derrick1964, Manton_et_Sutcliffe2004} and \cite{Andrews_et_al2010} which considered instabilities in spherically symmetric solutions. The paper \cite{Hinterbichler_et_al2010} showed that the N-galileons are the goldstone bosons of the spontaneous symmetry breaking of the $N+D$ dimensional diffeomorphism group in forming a codimension $N$ brane of $D$ dimensions. This property of the multigalileons was also mentioned in \cite{Padilla_et_al2010a} where it was noted that the bi-galileons were the goldstones from cascading gravity from 6 dimensions.

  The fact that the galileon is a natural artifact of a five dimensional theory was further reviewed in \cite{de_Rham_et_Tolley2010} which found in this a unification of the DBI \cite{Silverstein_et_Tong2004, Alishahiha_et_al2004, dbi,multi,Tong2004}  and galileon theories. The two theories turn out to be different nonlinear realisations of the 5D diffeomorphisms on a 4D brane. A generalisation of this theory was formulated and explored in \cite{Goon_et_al2010}.

  The connection between massive gravity and higher dimensional theories are well known, and in the case of the galileon the connection to massive spin 2 fields was shown in \cite{de_Rham_et_Gabadadze2010}.

  Many other generalisations of the galileon model have also been explored. In \cite{deFelice_et_Tsujikawa2010b} a generalisation where $\pi \to f(\pi)$ was considered in a cosmological context. Some of these generalisations break the galilean invariance. For instance a generalisation to the coefficients of the different galileon Lagrangian terms being functions of the galileon field was considered for up to third order galileons in \cite{Kobayashi2010}, \cite{Silva_et_Koyama2009}, \cite{Kobayashi_et_al2010a} and \cite{deFelice_et_Tsujikawa2010c}. Up to third order galileons have also been generalised to theories with coefficients not only dependent on the galileon field itself, but also on its kinetic term $X = - g^{\mu\nu}\partial_\mu\pi\partial_\nu\pi/2$ in \cite{deFelice_Mukohyama_et_Tsujikawa2010} which considered perturbation equations in such a generalised scenario, in \cite{Deffayet_et_al2010} where cosmological evolution was reviewed and in \cite{Kobayashi_et_al2010b} which considered inflation.

  Though the original formulation of the galileon was proposed as a dark energy alternative \cite{Nicolis_et_al2008}, the model has recently been studied as an alternative to inflation. In particular its NEC-violating  properties have attracted attention in this context as this can give rise to a galileon genesis from flat space \cite{Creminelli_et_al2010} yielding a theory with no Big Bang. The NEC-violation was also found in the generalised theory of \cite{Kobayashi_et_al2010b} in the inflationary scenario. In \cite{Burrage_et_al2010} a covariant version of the galileon is used to drive inflation, motivated by the fact that the constant shift symmetry is already only mildly broken in the traditional slow-roll inflation, and hence the galileon is a natural and useful place to look for generalised inflation.

  We want to take the model back to its basic formulation in \cite{Nicolis_et_al2008}, where we feel that a simple self-accelerating solution has yet to be properly explored. In this paper we find from the theoretical constraints in \cite{Nicolis_et_al2008} and \cite{Nicolis_et_al2009} that a third order galileon theory with a first order term, a so-called \emph{tadpole}, seems to be allowed, stable and yielding selfaccelerating solutions in a certain regime of the parameter space. We explore this simplest of the galileon selfaccelerating solution and its background evolution, as we feel that a thorough treatment of this theory seems the natural place to go first from the results of the original papers \cite{Nicolis_et_al2008} and \cite{Nicolis_et_al2009}. A treatment of the perturbations for this model will be postponed for a future publication.

  \section{The galileon Lagrangian}
  Before we start the treatment of our particular galileon model, we would like to recapitulate the formulation of the galileon model given in \cite{Nicolis_et_al2008}.

  Looking at the action:
  \begin{equation}
    S = \int d^4x[\frac{1}{2}M_{PL}^2\sqrt{-\hat{g}}[\hat{R} + \frac{1}{2}\hat{h}_{\mu\nu}T^{\mu\nu} + \mathcal{L}_\pi + \pi T_{\mu}^{\mu}]
  \end{equation}
  where $M_{PL}$ is the four-dimensional reduced Planck mass, $\hat{R}$ is the Einstein frame Ricci scalar, $T^{\mu\nu}$ is the stress-energy tensor for the matter fields, $\frac{1}{2}\hat{h}_{\mu\nu}$ is the coupling between gravity and the matter fields compensating for the matter not feeling the same metric as the Einstein frame one and $\pi$ is the galileon. The equations of motion of the galileon Lagrangian is invariant under the galilean transformation
  \begin{equation}
    \pi \to \pi + c + b_{\mu}x^{\mu}
  \end{equation}

  In 4D there are 5 galilean invariant Lagrangians, one for each order of the galileon up to 5. Higher order terms involve only total derivative terms. Defining $\Pi \equiv \partial^\mu\partial_\nu\pi$, $[\ldots]$ to be the trace of the operator $\ldots$ and $\cdot$ to denote Lorentz-invariant contraction of indices, the five galilean invariant Lagrangians are:
  \begin{eqnarray}
    \mathcal{L}_1 &=& \pi\\
    \mathcal{L}_2 &=& -\frac{1}{2}\partial\pi\cdot\partial\pi\\
    \mathcal{L}_3 &=& -\frac{1}{2}[\Pi]\partial\pi \cdot\partial\pi\\
    \mathcal{L}_4 &=& -\frac{1}{4}\left([\Pi]^2\partial\pi \cdot\partial\pi - 2[\Pi]\partial\pi\cdot\Pi\cdot\partial\pi - [\Pi^2]\partial\pi\cdot\partial\pi + 2\partial\pi\cdot\Pi^2\cdot\partial\pi\right)\\
    \mathcal{L}_5 &=& -\frac{1}{5}\Big([\Pi]^3\partial\pi\cdot\partial\pi - 3[\Pi]^2\partial\pi\cdot\Pi\cdot\partial\pi - 3[\Pi][\Pi^2]\partial\pi\cdot\partial\pi\nonumber\\ 
    && + 6[\Pi]\partial\pi\cdot\Pi^2\cdot\partial\pi  + 2[\Pi^3]\partial\pi\cdot\partial\pi + 3[\Pi^2]\partial\pi\cdot\Pi\cdot\partial\pi \nonumber \\
    &&- 6\partial\pi\cdot\Pi^3\cdot\partial\pi\Big) \\
  \end{eqnarray}

  In general, then, the full galilean field Lagrangian takes the form
  \begin{equation}
    \mathcal{L}_\pi = \sum_{i = 1}^5 c_i \mathcal{L}_i
  \end{equation}

  For our selfaccelerating solution we want a stable deSitter attractor for a vanishing matter stress-energy tensor. Considering in this case the effect of the galileon on the cosmological evolution yields a configuration for the $\pi$ in terms of the wanted asymptotic value $H_0$ of the Hubble parameter in this case. Plugging this into the equation of motion for the galileon leads to the following condition on the coefficients of the Lagrangian\footnote{For the full derivation of this condition in detail see \cite{Nicolis_et_al2008}.}
  \begin{equation}\label{eq:deSitterGalConstraint}
    c_1 - 2c_2H_0^2 + 3c_3H_0^4 - 3c_4H_0^6 + \frac{3}{2}c_5H_0^8 = 0
  \end{equation}

  Now, we want the internal galilean symmetries and external Lorentz symmetries to be unbroken, so the perturbations should obey these symmetries, hence if our equations of motion for the galilean terms are given by 
  \begin{equation}
    \frac{\delta\mathcal{L}_p}{\delta\pi} = \sum_{i = 1}^5c_i\mathcal{E}_i
  \end{equation}
  then the perturbations to the equations of motion must be given by
  \begin{equation}
    \frac{\delta\mathcal{L}_p}{\delta\pi} = \sum_{i = 1}^5d_i\mathcal{E}_i
  \end{equation}
  where $d_i$ are the coefficients for the perturbations. To simplify the calculations we assume the coefficients $d_i$ to incorporate the zeroth order deSitter solution plus the perturbations. We can then find invertible relations between the $d_i$s and the $c_i$s for $i = 2\ldots 5$. 
  \begin{displaymath}
    \left[\begin{array}{l}
        c_2\\
        c_3\\
        c_4\\
        c_5
      \end{array}\right]
     = \left[\begin{array}{llll}
        1&3H_0^2&\frac{9}{2}H_0^4&3H_0^6\\
        0&1&3H_0^2&3H_0^4\\
        0&0&1&2H_0^2\\
        0&0&0&1
        \end{array}\right]
      \cdot \left[\begin{array}{l}
          d_2\\
          d_3\\
          d_4\\
          d_5
        \end{array}\right]
  \end{displaymath}
  
  $c_1$ is then simply given by the equation (\ref{eq:deSitterGalConstraint}).

  The inverse matrix becomes

  \begin{displaymath}
    \left[\begin{array}{l}
        d_2\\
        d_3\\
        d_4\\
        d_5
      \end{array}\right]
     = \left[\begin{array}{llll}
        1&-3H_0^2&\frac{9}{2}H_0^4&-3H_0^6\\
        0&1&-3H_0^2&3H_0^4\\
        0&0&1&-2H_0^2\\
        0&0&0&1
        \end{array}\right]
      \cdot \left[\begin{array}{l}
          c_2\\
          c_3\\
          c_4\\
          c_5
        \end{array}\right]
  \end{displaymath}
  We should now check what constraints must be put on these coefficients for the solution to admit a stable deSitter solution with spherical symmetric solutions about compact sources. These configurations should be ghost free. We also want the fluctuations to be subluminal.

  Combining all this demands the conditions for the $d_i$s are given by
  \begin{eqnarray}
          d_2 &>& 0\nonumber\\
          d_4 &\geq&  0\nonumber\\
          d_3 &\geq& \sqrt{\frac{3}{2}d_2d_4}\nonumber\\
          d_5 &<& 0\label{eq:dConditions}
    \end{eqnarray}

    We should study the cosmology of such a model where the null-energy condition is viable, and see whether it can reproduce results for $\Lambda$CDM cosmology in terms of background equations and perturbations.

    Actually the constraints above already ensure subluminality and no breaking of the NEC. In the coming subsection we check that this is actually true in the case considered in the article \cite{Nicolis_et_al2009}.

    \subsection{The case of $\mathcal{L}_1 = 0$}
    In the article \cite{Nicolis_et_al2009} they consider a deSitter solution with $c_1 = 0$ and check for the demand of no violation of the NEC conditions means that:
    \begin{equation}\label{eq:NoNECViolation}
      c_4 \leq \frac{2}{333}\frac{c_2}{H_0^4} + \frac{1}{111}\frac{c_3}{H_0^2}
    \end{equation}

    When $c_1 = 0$, then $c_5$ is completely determined from the deSitter criterion (\ref{eq:deSitterGalConstraint}). This means that we can rewrite our transformation matrices so that they only transform between indices $2,3,4$. The results are
   \begin{displaymath}
    \left[\begin{array}{l}
        c_2\\
        c_3\\
        c_4
      \end{array}\right]
     = \left[\begin{array}{llll}
        -3& -3H_0^2&-\frac{3}{2}H_0^4\\
        -\frac{4}{H_0^2}&-5&-3H_0^2\\
        -\frac{8}{3H_0^4}&-\frac{4}{H_0^2}&-3&
        \end{array}\right]
      \cdot \left[\begin{array}{l}
          d_2\\
          d_3\\
          d_4
        \end{array}\right]
  \end{displaymath}   
    
   \begin{displaymath}
    \left[\begin{array}{l}
        d_2\\
        d_3\\
        d_4
      \end{array}\right]
     = \left[\begin{array}{llll}
        -3& 3H_0^2&-\frac{3}{2}H_0^4\\
        \frac{4}{H_0^2}&-5&3H_0^2\\
        -\frac{8}{3H_0^4}&\frac{4}{H_0^2}&-3&
        \end{array}\right]
      \cdot \left[\begin{array}{l}
          c_2\\
          c_3\\
          c_4
        \end{array}\right]
  \end{displaymath}   

  By using the transformation matrices above to turn the condition on $c$s given in equation (\ref{eq:NoNECViolation}) into a condition on the $d$s, we realise that the conditions given in equation (\ref{eq:dConditions}) are sufficient to prohibit the breaking of the NEC.

  \section{A possible Lagrangian for the galilean modified gravity with selfaccelerated solutions}

  Since the demands on the coefficients already ensures a deSitter solution, we should look at the early evolution and the evolution of perturbations in this theory. Since the assumptions also imply stability of the deSitter solution we should only check the stability in the more general non-deSitter regime.

  However, constraints from spherical sources compels us to choose $d_4 = d_5 + c_4 = c_5 = 0$.\footnote{See \cite{Nicolis_et_al2008}} Hence we are left only with $c_2$ and $c_3$. The constraint in equation (\ref{eq:dConditions}) becomes
  \begin{eqnarray}
    c_1 - 2c_2H_0^2 +3c_3H_0^4 = c_1 - 2d_2H_0^2 - 3d_3H_0^4 = 0
  \end{eqnarray}
  and requires $c_1$ to be larger than zero for the constraints on the $d$s to hold. Hence we get a Lagrangian:

  \begin{equation}
    L_\pi = c_1 \pi - \frac{c_2}{2}(\partial\pi)^2 - \frac{c_3}{2}\Box\pi(\partial \pi)^2
  \end{equation}
  where 
  \begin{eqnarray}\label{eq:ConstraintsForFullModel}
    c_3 &\geq& 0\nonumber\\
    c_2 &>& 3H_0^2c_3\nonumber\\
    c_1 &=& 2c_2H_0^2 - 3H_0^4c_3
  \end{eqnarray}
  where $H_0$ is the value of the Hubble rate in the deSitter attractor solution. In the later numerical computations we set this value to the present day Hubble rate.

  We choose to work in a Jordan frame where matter is minimally coupled to the galileon, thus considering the galileon a modification to gravity as was eventually also done in the formulation made by \cite{Nicolis_et_al2008}. Our final action becomes
  \begin{equation}\label{eq:actionJordan}
    S = \int d^4x\sqrt{-g}\left(\frac{M_{\mathrm{PL}}^2}{2}e^{-2\pi}R + c_1\pi - \frac{c_2}{2}\left(\partial\pi\right)^2 - \frac{c_3}{2}\Box\pi(\partial \pi)^2\right) + S_{\mathrm{matter}}
  \end{equation}
where $S_{\mathrm{matter}}$ is the action of other fields like photons, neutrinos, baryons and dark matter. 

Because of the coupling between the galileon and gravity the resulting Einstein equations and equation of motion for the galileon field are no longer galilean invariant in a non-flat background. However, since the Jordan frame has been chosen a coupling between gravity and the galileon must be supplied. This particular choice makes the action reduce to the ordinary Einstein-Hilbert action in the limit of no galileon fields, whilst keeping the galilean symmetry in flat space. Hence this choice for a coupling to gravity is natural. It is also the same non-minimal coupling as chosen in \cite{Nicolis_et_al2008} so all their stability and selfacceleration conditions hold for this model.

  From this and reviewing the result of \cite{Chow_et_Khoury} we can find the equations of motion. For $\pi$ it is:
  \begin{equation}\label{eq:eomPi}
    c_1 + c_2\Box\pi + c_3 \left(\left(\Box \pi\right)^2 - \left(\nabla_\mu\nabla_\nu\pi\right)^2 - R^{\mu\nu}\nabla_\mu\pi\nabla_\nu\pi\right) = M_{\mathrm{PL}}^2Re^{-2\pi}
  \end{equation}
  Using the result from \cite{Chow_et_Khoury} combined with the derivation in \cite{Pimentel1989} the Einstein equations become
  \begin{eqnarray}
    e^{-2\pi}M_{\mathrm{PL}}^2G_{\mu\nu} &=& T_{\mu\nu} + M_{\mathrm{PL}}^2\left(\nabla_\mu\nabla_\nu - g_{\mu\nu}\Box\right)e^{-2\pi} + \nonumber \\
    && c_1\pi g_{\mu\nu} + c_2\left(\nabla_\mu\pi\nabla_\nu\pi - \frac{1}{2}g_{\mu\nu}\nabla_\alpha\pi\nabla^\alpha\pi\right) + \nonumber\\
    &&\frac{c_3}{2}\left(2\nabla_\mu\pi\nabla_\nu\pi\Box\pi + g_{\mu\nu}\nabla_\alpha\pi\nabla^\alpha\left(\partial\pi\right)^2 - 2\nabla_{(\mu}\pi\nabla_{\nu)}\left(\partial\pi\right)^2\right)\nonumber\\
    \label{eq:Einstein_eqs_general}
  \end{eqnarray}

  This should enable us to study background cosmology and eventually also perturbation theory in this modified version of gravity.

  \subsection{Equations of motion in a homogeneous and isotropic Universe}

  To simplify calculations and the considerations of the equations of motion derived above we start by assuming isotropy and homogeneity. We also assume the Universe to be spatially flat. This means that we are working in a flat FRW metric background. This derivation is very similar to the one presented in section 4 of \cite{Chow_et_Khoury}.

  In the flat FRW metric background, the equation of motion for $\pi$ in equation (\ref{eq:eomPi}) now becomes:
  \begin{eqnarray}
  M_{\mathrm{PL}}^2 6(\frac{\ddot{a}}{a} + H^2)e^{-2\pi} &=& c_1 - c_2\left(\ddot{\pi} + 3H\dot{\pi}\right) + \nonumber\\ 
  && 3c_3\left(\frac{\partial}{\partial t}\left(H\dot{\pi}^2\right) + 3H^2\dot{\pi}^2\right)\label{eq:PhiEOM}
   \end{eqnarray}
   where we see that the term proportional to $c_3$ is equivalent to the term found in \cite{Chow_et_Khoury}.

   The $00$ component of the Einstein equation gives the equivalent of the modified Friedmann equation
   \begin{equation}\label{eq:genFriedmann}
     3e^{-2\pi}M_{\mathrm{PL}}^2H^2 = \rho  - c_1\pi + \frac{c_2}{2}\dot{\pi}^2  + 6 H \dot{\pi}M_{\mathrm{PL}}^2e^{-2\pi} - 3Hc_3\dot{\pi}^3
   \end{equation}

   The combination $-\frac{1}{6}\left(00 + \frac{3}{a^2}ii\right)$ yields the Raychaudhuri equation
   \begin{eqnarray}
     M_{\mathrm{PL}}^2e^{-2\pi}\frac{\ddot{a}}{a} &=& -\frac{1}{6}\left(\rho + 3P\right) - \frac{1}{3}\left(c_1\pi + c_2 \dot{\pi}^2\right) + \frac{c_3}{2}\dot{\pi}^2\left( H\dot{\pi} - \ddot{\pi}\right) \nonumber \\
     && + M_{\mathrm{PL}}^2e^{-2\pi}\left(H\dot{\pi} + \ddot{\pi} - 2\dot{\pi}^2\right)\label{eq:genRaychaudhuri} 
   \end{eqnarray}
   which has a slight discrepancy of $e^{-2\pi}H\dot{\pi}M_{\mathrm{PL}}^2$ with respect to the one found in \cite{Chow_et_Khoury} which seems to be due to a difference in calculations of covariant derivatives.

   As in \cite{Chow_et_Khoury} we can find an effective energy density $\rho_\pi$ and pressure $P_{\pi}$ for the $\pi$ field.
   \begin{eqnarray}
     \rho_{\pi} &=& -c_1\pi + \frac{c_2}{2}\dot{\pi}^2 + 6H\dot{\pi}M_{\mathrm{PL}}^2e^{-2\pi} - 3Hc_3\dot{\pi}^3\\
     P_{\pi} &=& c_1\pi + \frac{c_2}{2}\dot{\pi}^2 + c_3\ddot{\pi}\dot{\pi}^2 - 4M_{\mathrm{PL}}^2H\dot{\pi}e^{-2\pi} \nonumber \\
     &&- 2M_{\mathrm{PL}}^2\ddot{\pi}e^{-2\pi} + 4M_{\mathrm{PL}}^2\dot{\pi}^2e^{-2\pi}
   \end{eqnarray}
   Again the part of $P_\pi$ resulting from $\mathcal{L}_3$ is not exactly the same as found in \cite{Chow_et_Khoury}. The same discrepancy, presumably caused by the same calculations difference, is still present.

   \section{Numerical simulation of the background equations}

   Before we start the numerical simulations of the background equations (\ref{eq:PhiEOM}), (\ref{eq:genFriedmann}) and (\ref{eq:genRaychaudhuri}) it is useful to rewrite the derivatives with respect to $t$ in terms of a new variable $x = \ln a$. Since $\frac{d}{dt} = H\frac{d}{dx}$ we get for the equation of motion of the galileon:
   \begin{eqnarray}
     M_{\mathrm{PL}}^26H\left(H' + 2H\right)e^{-2\pi} &=& c_1 - c_2H\left(H'\pi' + H\pi'' + 3H\pi'\right) + \nonumber\\
     &&3c_3\left(H\frac{\partial}{\partial x}\left(H^3\left(\pi'\right)^2\right) + 3H^4\left(\pi'\right)^2\right)\nonumber\\
     &=& c_1 - c_2H\left(H'\pi' + H\pi'' + 3H\pi'\right) + \nonumber\\
     && 3H^3c_3\pi'\left(3H'\pi' + 2H\pi'' + 3H\pi'\right)\nonumber \\\label{eq:galPiEOMX}
   \end{eqnarray}
   where $'$ denotes a derivative with respect to $x$.

   The Friedmann and Raychaudhuri equations come out as
   \begin{equation}\label{eq:galFriedmannX}
     3e^{-2\pi}M_{\mathrm{PL}}^2H^2 = \rho  - c_1\pi + \frac{c_2}{2}H^2\left(\pi'\right)^2  + 6 H^2\pi'M_{\mathrm{PL}}^2e^{-2\pi} - 3H^4c_3\left(\pi'\right)^3
   \end{equation}
   and
   \begin{eqnarray}
      M_{\mathrm{PL}}^2e^{-2\pi}H\left(H' + H\right) &=& -\frac{1}{6}\left(\rho + 3P\right) - \frac{1}{3}\left(c_1\pi + c_2 H^2\left(\pi'\right)^2\right) + \nonumber \\
      && \frac{c_3}{2}H^3\left(\pi'\right)^2\left( H\pi' - H'\pi' - H\pi''\right) \nonumber \\
     && + HM_{\mathrm{PL}}^2e^{-2\pi}\left(H \pi' +H'\pi' + H\pi'' - 2H\left(\pi'\right)^2\right)\nonumber\\ \label{eq:galRaychaudhuriX}
   \end{eqnarray}

   We now take advantage of the nice property of the Jordan frame that matter is not coupled to the galileon field. This means that the equations that determine the development of the matter or radiation from their equations of state are not altered, so $\rho_m = \Omega_ma^{-3} = \Omega_m\exp\left(-3x\right)$ and $\rho_\gamma = \Omega_\gamma a^{-4} = \Omega_{\gamma}\exp\left(-4x\right)$, $\rho = \rho_m + \rho_\gamma$ and $\rho + 3P = \rho_m + 2\rho_\gamma$.

   We use the above  equations to make a set of three coupled differential equations to solve numerically in $H$, $\pi$ and $\pi'$. First we use the equation:
   \begin{equation}
     \frac{d\pi}{dx} = \pi'
   \end{equation}
   thereby transforming the set of second order equations (\ref{eq:galPiEOMX}-\ref{eq:galRaychaudhuriX}) in $\pi$ into a set of first order equations in $\pi$, $\pi'$ and $H$. Then we need an equation for $\pi''$ and one for $H'$. However, we need at least one of these not to depend on the other.

   We enter the result from the first Friedmann equation into the Raychaudhuri equation, to simplify slightly. The result is:
   \begin{eqnarray}
 M_{\mathrm{PL}}^2e^{-2\pi}H\left(H' + \frac{3}{2}H\right) &=& -\frac{1}{2}P - \frac{1}{2}c_1\pi - \frac{1}{4}c_2 H^2\left(\pi'\right)^2 -\frac{c_3}{2}H^3\left(\pi'\right)^2\left(H'\pi' + H\pi''\right) \nonumber \\
     && + HM_{\mathrm{PL}}^2e^{-2\pi}\left(2H \pi' +H'\pi' + H\pi'' - 2H\left(\pi'\right)^2\right)\nonumber\\ \label{eq:galRaychaudhuriXPlusFriedman}
   \end{eqnarray}

   Then using (\ref{eq:galRaychaudhuriXPlusFriedman}) to solve for the second derivative of the galileon field we get:
   \begin{equation}
     \pi'' = \gamma + \delta H'
   \end{equation}
   where
   \begin{eqnarray}
     \gamma &=& \frac{M_{\mathrm{PL}}^2H^2e^{-2\pi}\left(\frac{3}{2} - 2\pi' + 2\left(\pi'\right)^2\right) + \frac{1}{2}P + \frac{1}{2}c_1\pi + \frac{1}{4}c_2H^2\left(\pi'\right)^2}{H^2\left(M_{\mathrm{PL}}^2e^{-2\pi} - \frac{c_3}{2}H^2\left(\pi'\right)^2\right)}\nonumber\\
     \delta &=&\frac{M_{\mathrm{PL}}^2e^{-2\pi}\left(1 - \pi'\right) + \frac{c_3}{2}H^2\left(\pi'\right)^3}{H\left(M_{\mathrm{PL}}^2e^{-2\pi} - \frac{c_3}{2}H^2\left(\pi'\right)^2\right)}\label{eq:gamma_and_delta_general}
   \end{eqnarray}
   
   Plugging this into equation (\ref{eq:galPiEOMX}) we get that
   \begin{equation}\label{eq:H_for_num_gen}
     H' = \frac{c_1 - c_2H^2\left(\gamma + 3\pi'\right) + 3H^4c_3\pi'\left(2\gamma + 3\pi'\right) - 12H^2M_{\mathrm{PL}}^2e^{-2\pi}}{6HM_{\mathrm{PL}}^2e^{-2\pi} + c_2H\left(\pi'+H\delta\right)- 3c_3H^3\pi'\left(3\pi' + 2H\delta\right)}
   \end{equation}

   For reference it is nice to have the formulas for the galileon density and pressure given in terms of the $'$ derivatives:

   \begin{eqnarray}
     \rho_{\pi} &=& -c_1\pi + \frac{c_2}{2}H^2\left(\pi'\right)^2 + 6H^2\pi'M_{\mathrm{PL}}^2e^{-2\pi} - 3H^4c_3\left(\pi'\right)^3\\
     P_{\pi} &=& c_1\pi + \frac{c_2}{2}H^2\left(\pi'\right)^2 + c_3H^3\left(\pi'\right)^2\left(H'\pi' + H\pi''\right) - 4M_{\mathrm{PL}}^2H^2\pi'e^{-2\pi} \nonumber \\
     &&- 2M_{\mathrm{PL}}^2He^{-2\pi}\left(H'\pi' + H\pi''\right) + 4M_{\mathrm{PL}}^2H^2\left(\pi'\right)^2e^{-2\pi}
   \end{eqnarray}

   \section{Results in different particular regimes}
   Before moving on to our full model it is interesting to review and study the behaviour in different regimes of special values for the coefficients. Some of these may be interesting for their simplicity. Others may be comparable to theories considered elsewhere. Together they provide useful intuition for our interpretation of the evolution of the full model.

   \subsection{All coefficients zero}\label{sec:all_coeffs_0}
   If we set $c_1 = c_2 = c_3 = 0$, the galileon field will still be present and dynamical, and will have a special effect on the evolution of the Universe. We can see this directly from writing out the equations in this case:

   The galileon equation of motion gives:
   \begin{equation}
     \frac{\ddot{a}}{a} + H^2 = 0
   \end{equation}
   Adding the Friedmann equation and Einstein equations together and writing $\rho_{\mathrm{tot}} = \rho + \rho_\pi$, $P_{\mathrm{tot}} = P + P_\pi$, we get:
   \begin{equation}
     \frac{\ddot{a}}{a} + H^2 = -\frac{1}{6}\left(\rho_{\mathrm{tot}} + 3P_{\mathrm{tot}}\right) + \frac{1}{3}\rho_{\mathrm{tot}}\mathrm{,}
   \end{equation}
   which leads to a demand for an effective total equation of state at all times of 
   \begin{equation}
     w_{\mathrm{eff}} \equiv \frac{P_{\mathrm{tot}}}{\rho_{\mathrm{tot}}} = \frac{1}{3}\mathrm{.}
   \end{equation}
   This means that the Universe appears radiation dominated at all times. Numerical results show this, as can be seen from the evolution of $H$ shown in figure \ref{fig:H_all_coeffs0}.
\
\FIGURE{
      \includegraphics[width=0.6\textwidth]{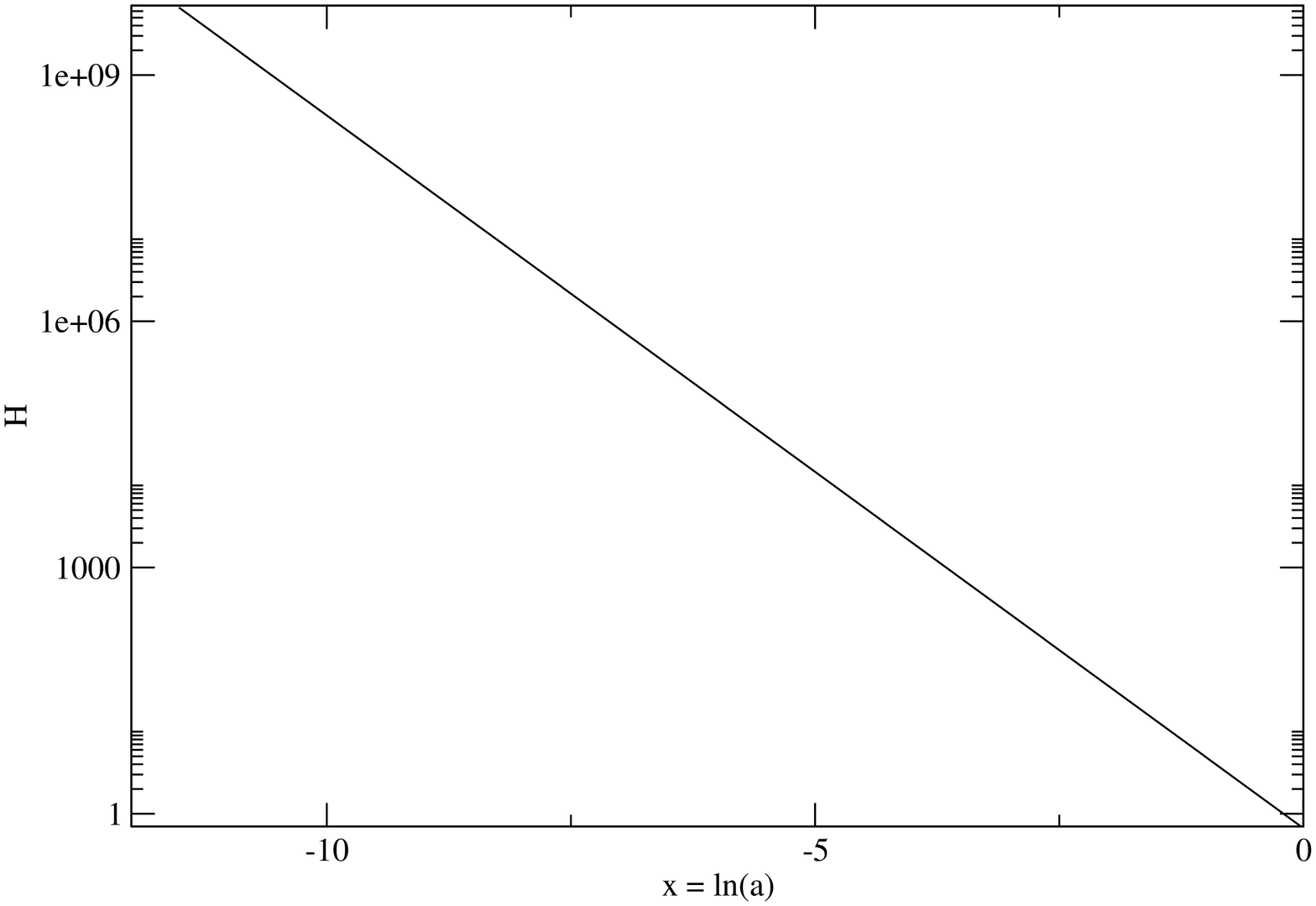}
      \caption{ Plot of the evolution of the Hubble parameter $H$ in the galileon model with all coefficients $c_1 = c_2 = c_3 = 0$. Note that $H$ evolves as if always in a radiation dominated Universe.}
      \label{fig:H_all_coeffs0}
}

  \subsection{Tadpole only, $c_2 = c_3 = 0$, non zero $c_1$.}\label{sec:Tadpole}
  The case of $c_1$ being the only non-zero constant greatly resembles the case considered above in section \ref{sec:all_coeffs_0} with no coefficients. However we now have a linear potential making the theory start off in such a way that the evolution starts not like a radiation dominated one, but one defined by the constant $c_1$ if the galileon starts out very small. We see this from the equation of state of the galileon in this case:
  \begin{equation}\label{eq:HprimeTadpole}
\left(\frac{\ddot{a}}{a}\right) + H^2 =    e^{2\pi}\frac{c_1}{6 M_{\mathrm{PL}}^2}
  \end{equation}
  As the galileon grows the term on the right will grow leading eventually to accelerated expansion. If the galileon field on the other hand has negative values, the evolution will move towards the radiation like one as the absolute value of the field becomes larger.

  Adding together the Friedmann and Raychaudhuri equations we get that
  \begin{equation}
    c_1 = \rho_{\mathrm{tot}}\left(1 - 3w_{\mathrm{eff}}\right)
  \end{equation}
  We note that as $c_1 \neq 0$ the theory can only approach a radiation dominated like epoch asymptotically as $\rho_{\mathrm{tot}}$ gets very large.

  In the tadpole only case the equations for the numerical simulations become:
  \begin{eqnarray}
     \gamma &=& \frac{3}{2} - 2\pi' + 2\left(\pi'\right)^2 + \frac{P + c_1\pi}{2H^2M_{\mathrm{PL}}^2e^{-2\pi}}\nonumber\\
     \delta &=&\frac{1 - \pi'}{H}\label{eq:gamma_and_delta_tadpole}
   \end{eqnarray}
   and $H'$ is simply given by equation (\ref{eq:HprimeTadpole}).
   

   For very small values of $c_1$ we of course retain results close to those in the case discussed in section \ref{sec:all_coeffs_0}. For larger values of $c_1$ the results are shown in figure \ref{fig:tadpole_H} for the evolution of $H$ and in figure \ref{fig:tadpole_w} for the evolution of the equation of state parameter.

   From the two plots we see that in the beginning the Universe behaves as if radiation dominated, or very nearly so. This is because early on $\rho_{\mathrm{tot}}$ is very large so that we approach the asymptotic value for the equation of state parameter $w_{\mathrm{eff}} = 1/3$. As the evolution goes on it deviates from this and starts showing a smaller equation of state parameter and a smaller deceleration. This shift happens earlier, giving smaller values for the equation of state parameter and slower deceleration the larger the value of $c_1$. For sufficiently large values of $c_1$ a period of acceleration can even be reached. However, after experiencing a minimum value of the equation of state parameter or deceleration, the equation of state parameter starts growing again eventually asymptotically towards an evolution with $0<w_{\mathrm{eff}}<1/3$. We see that by making $c_1$ large enough we can get a value for $H$ today equal to the one we find from current observations\footnote{See for instance \cite{Komatsu_et_al2010}.}. However, in these cases the accelerated expansion has already taken place and passed, the Universe now being in a decelerated state. We also see that most of the evolution history will change considerably in this case. 

\FIGURE{
      \includegraphics[width=0.6\textwidth]{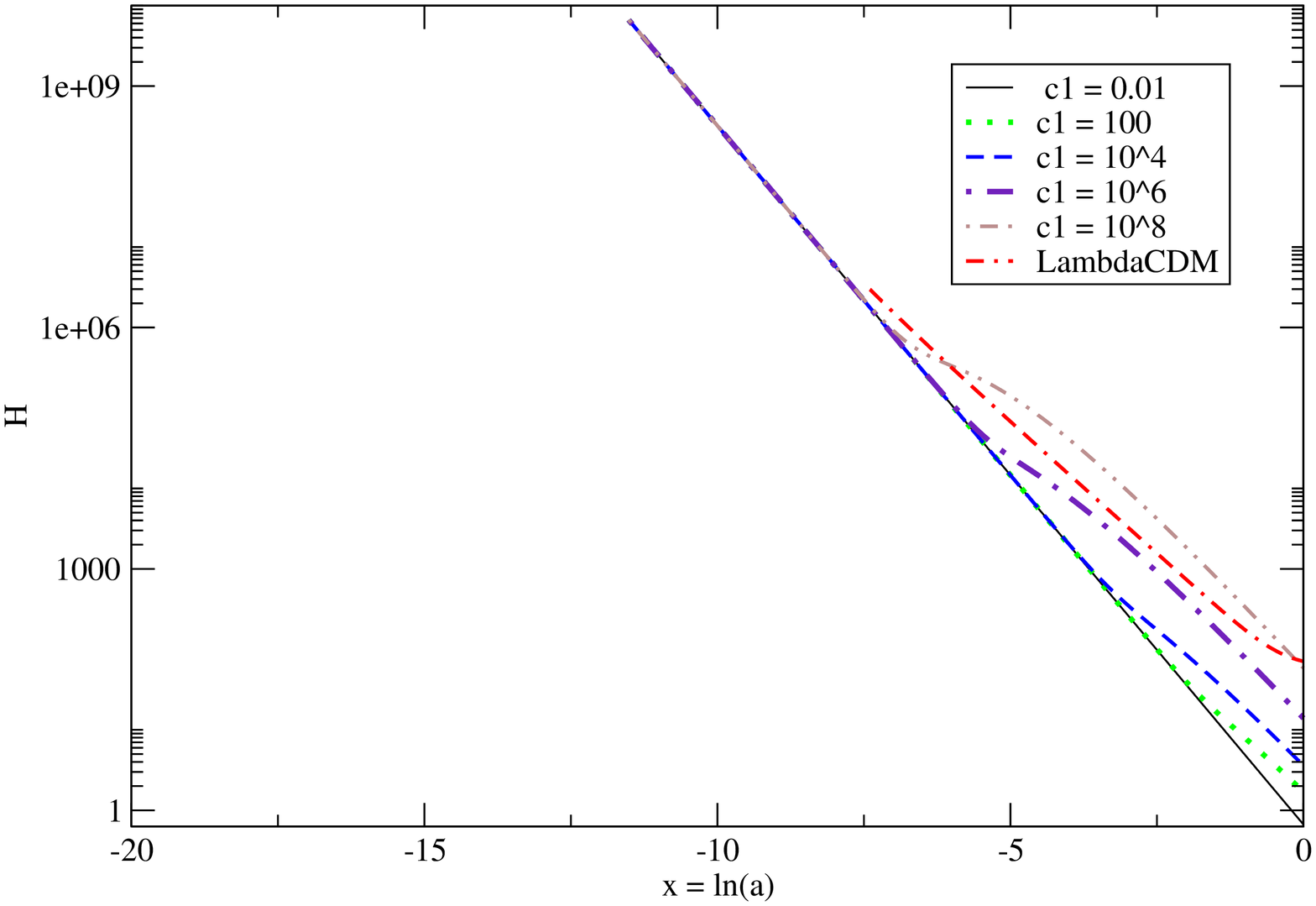}
      \caption{Plot of the evolution of the Hubble parameter $H$ in the galileon model with coefficients $c_2 = c_3 = 0$ and nozero $c_1$. The evolution of $H$ is shown for five different values of $c_1$: $c_1 = 0.01M_{\mathrm{PL}}^2H_0^2$, $c_1 = 100M_{\mathrm{PL}}^2H_0^2$, $c_1 = 10^4M_{\mathrm{PL}}^2H_0^2$, $c_1 = 10^6M_{\mathrm{PL}}^2H_0^2$ and $c_1 = 10^8M_{\mathrm{PL}}^2H_0^2$. The evolution of $H$ for the usual $\Lambda$CDM model has also been plotted alongside the others.}
      \label{fig:tadpole_H}}

      \FIGURE{
      \includegraphics[width=0.6\textwidth]{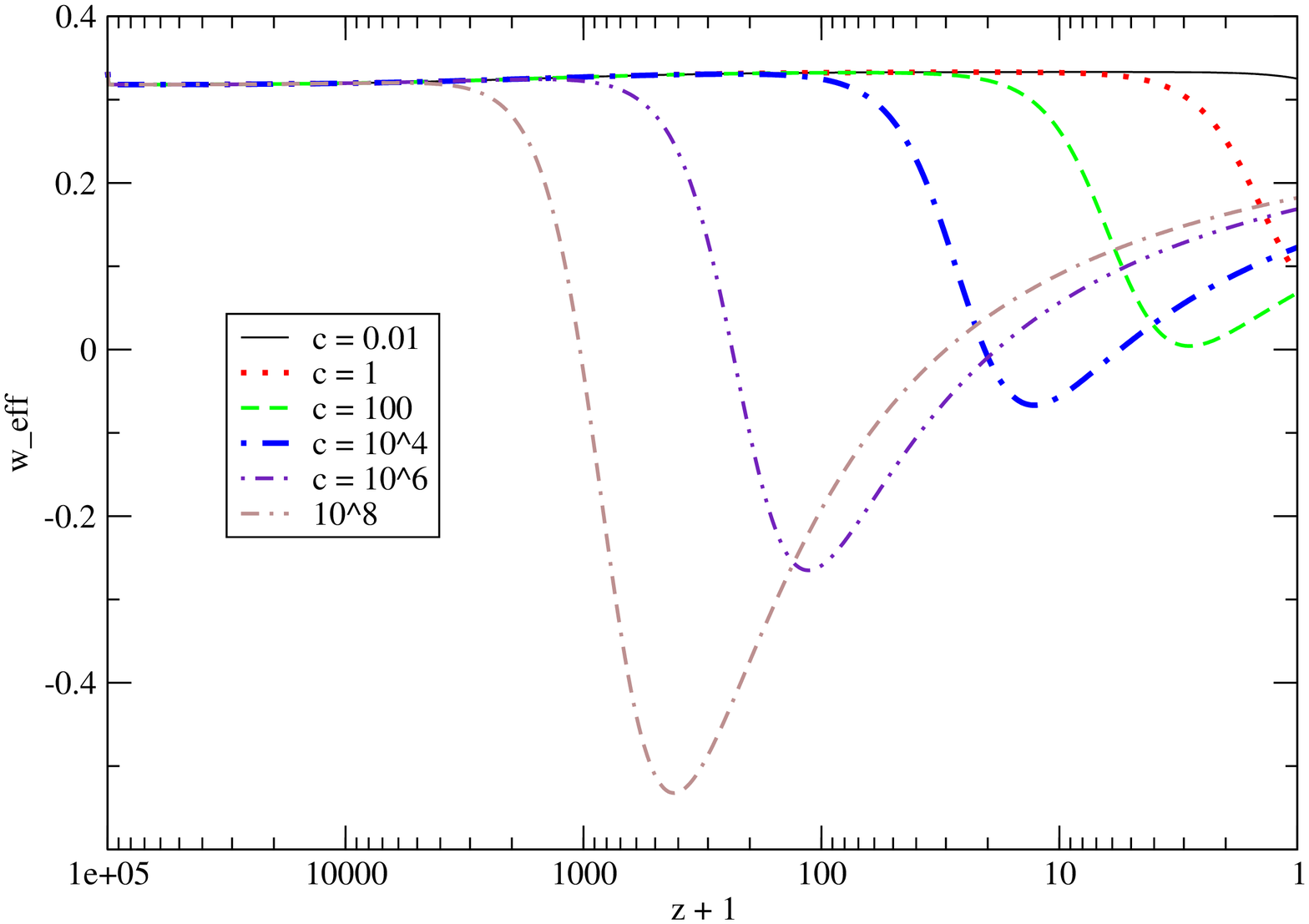}
      \caption{Plot of the evolution of the total equation of state parameter $w_{\mathrm{eff}}$ in the galileon model with coefficients $c_2 = c_3 = 0$, $c_1$ nonzero. The evolution is shown for six different values of $c_1$: $c_1 = 0.01M_{\mathrm{PL}}^2H_0^2$, $c_1 = M_{\mathrm{PL}}^2H_0^2$, $c_1 = 100M_{\mathrm{PL}}^2H_0^2$, $c_1 = 10^4M_{\mathrm{PL}}^2H_0^2$, $c_1 = 10^6M_{\mathrm{PL}}^2H_0^2$ and $c_1 = 10^8M_{\mathrm{PL}}^2H_0^2$.}
      \label{fig:tadpole_w}
  }

  Considering specifically the two cases of largest $c_1$, $c_1 = 10^6M_{\mathrm{PL}}^2H_0^2$ and $c_1 = 10^8M_{\mathrm{PL}}^2H_0^2$ since these have the most interesting evolution, we plot the values of the different densities and fractional densities in figures \ref{fig:tadpole_Om_106}, \ref{fig:tadpole_Om_108}, \ref{fig:tadpole_rho_106} and \ref{fig:tadpole_rho_108}.

\FIGURE{
      \includegraphics[width=0.6\textwidth]{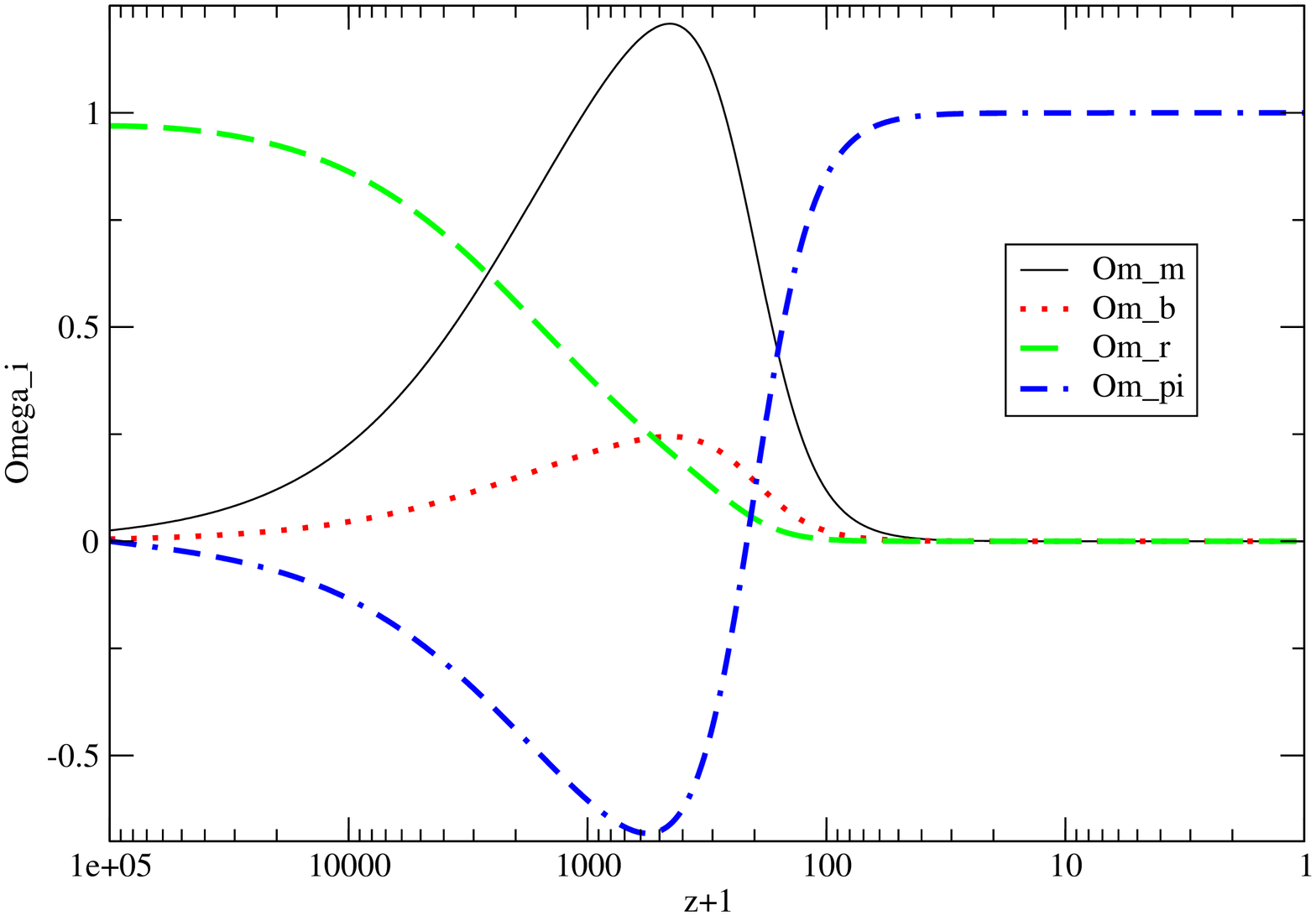}
      \caption{Plot of the evolution of the fractional densities $\Omega_i$ in the galileon model with coefficients $c_2 = c_3 = 0$, $c_1 = 10^6M_{\mathrm{PL}}^2H_0^2$.}
      \label{fig:tadpole_Om_106}
  }

\FIGURE{
      \includegraphics[width=0.6\textwidth]{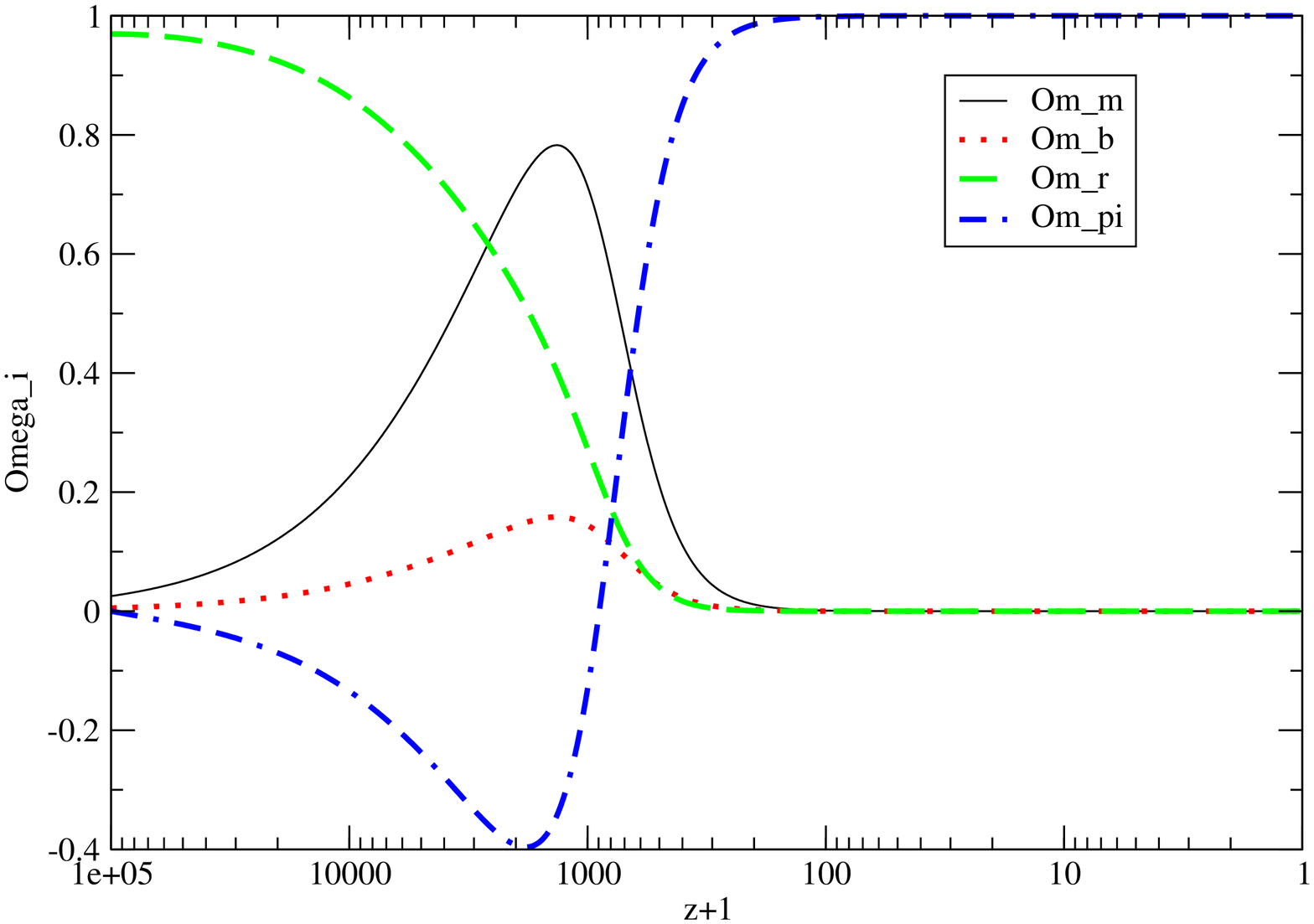}
      \caption{Plot of the evolution of the fractional densities $\Omega_i$ in the galileon model with coefficients $c_2 = c_3 = 0$, $c_1 = 10^8M_{\mathrm{PL}}^2H_0^2$.}
      \label{fig:tadpole_Om_108}
  }

\FIGURE{
      \includegraphics[width=0.6\textwidth]{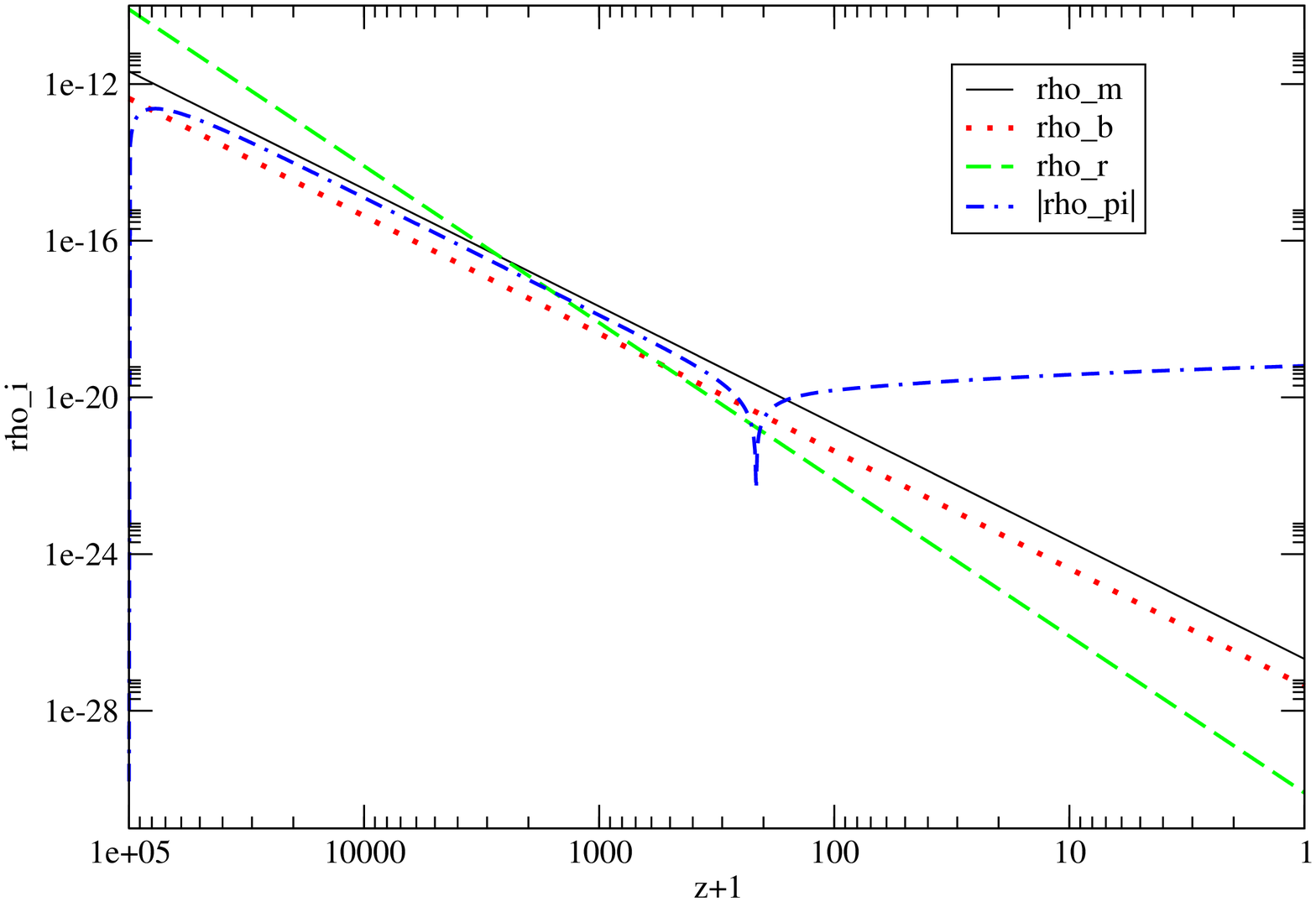}
      \caption{Plot of the evolution of the densities $\rho_i$ in the galileon model with coefficients $c_2 = c_3 = 0$, $c_1 = 10^6M_{\mathrm{PL}}^2H_0^2$. Since the galileon density starts out negative its absolute value $|\rho_{\pi}|$ is shown in the plot.}
      \label{fig:tadpole_rho_106}
    }

\FIGURE{
      \includegraphics[width=0.6\textwidth]{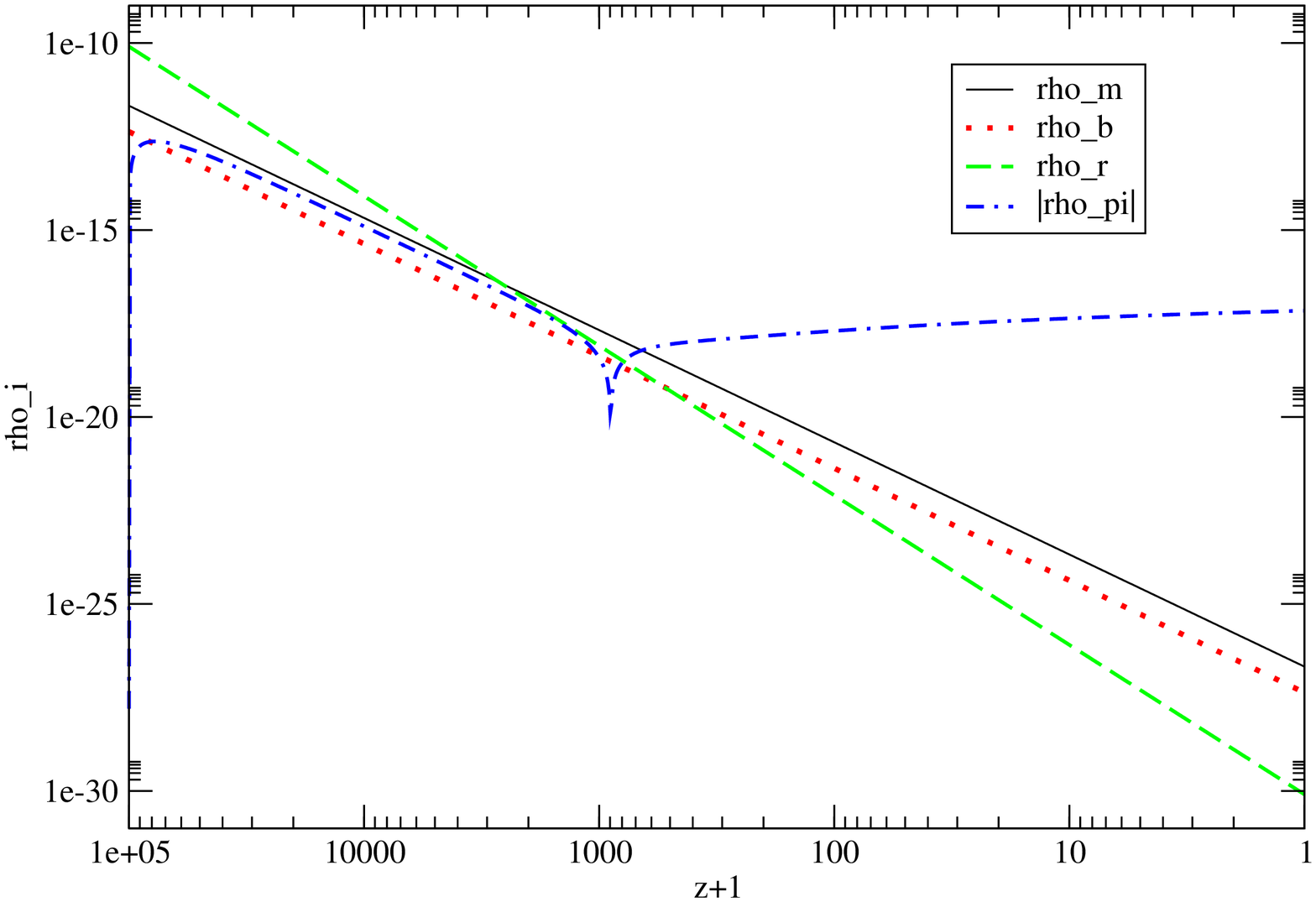}
      \caption{Plot of the evolution of the densities $\rho_i$ in the galileon model with coefficients $c_2 = c_3 = 0$, $c_1 = 10^8M_{\mathrm{PL}}^2H_0^2$. Since the galileon density starts out negative its absolute value $|\rho_{\pi}|$ is shown in the plot.}
      \label{fig:tadpole_rho_108}
  }

  \subsection{$c_1 = c_3 = 0$, non zero $c_2$ - the Brans-Dicke type theory}\label{sec:Brans-Dicke}
  The case of only $c_2$ different from zero is in fact a type of Brans-Dicke theory \cite{Brans_et_Dicke1961}. We can see this from starting with the generic Brans-Dicke action:
  \begin{equation}
    S_{\mathrm{BD}}  = \int d^4x\sqrt{-g}\left[\phi R - \frac{\omega(\phi)}{\phi}g^{\mu\nu}\partial_{\mu}\phi\partial_{\nu}\phi\right]
  \end{equation}

  To get this into a form similar to our own we realise that we must set:
  \begin{equation}
    \phi = \frac{M_{\mathrm{PL}}}{2}^2e^{-2\pi}
  \end{equation}
  and inserting this we see that for the second term to correspond to our $c_2$ term we must have:
  \begin{equation}
    \omega(\phi) = \frac{c_2}{8\phi}
  \end{equation}
  I.e. a theory with only the $c_2$ term present is equivalent to a Brans-Dicke theory with the Brans-Dicke parameter $\omega(\phi) \propto \frac{1}{\phi}$.

  The background equations we need to simulate now have:

     \begin{eqnarray}
     \gamma &=& \frac{3}{2} - 2\pi' + 2\left(\pi'\right)^2 + \frac{2P + c_2H^2\left(\pi'\right)^2}{4H^2M_{\mathrm{PL}}^2e^{-2\pi}}\nonumber\\
     \delta &=&\frac{1 - \pi'}{H}\label{eq:gamma_and_delta_BD}
   \end{eqnarray}
   
   Plugging this into equation (\ref{eq:galPiEOMX}) we get that
   \begin{equation}\label{eq:H_for_num_BD}
     H' = -\frac{c_2H^2\left(\gamma + 3\pi'\right) + 12H^2M_{\mathrm{PL}}^2e^{-2\pi}}{6HM_{\mathrm{PL}}^2e^{-2\pi} + c_2H\left(\pi'+H\delta\right)}
   \end{equation}

   The results for the evolution of the Hubble parameter $H$ and the effective equation of state parameter for different values of $c_2$ is shown in figures \ref{fig:BD_H} and \ref{fig:BD_w} respectively.

      \FIGURE{
      \includegraphics[width=0.6\textwidth]{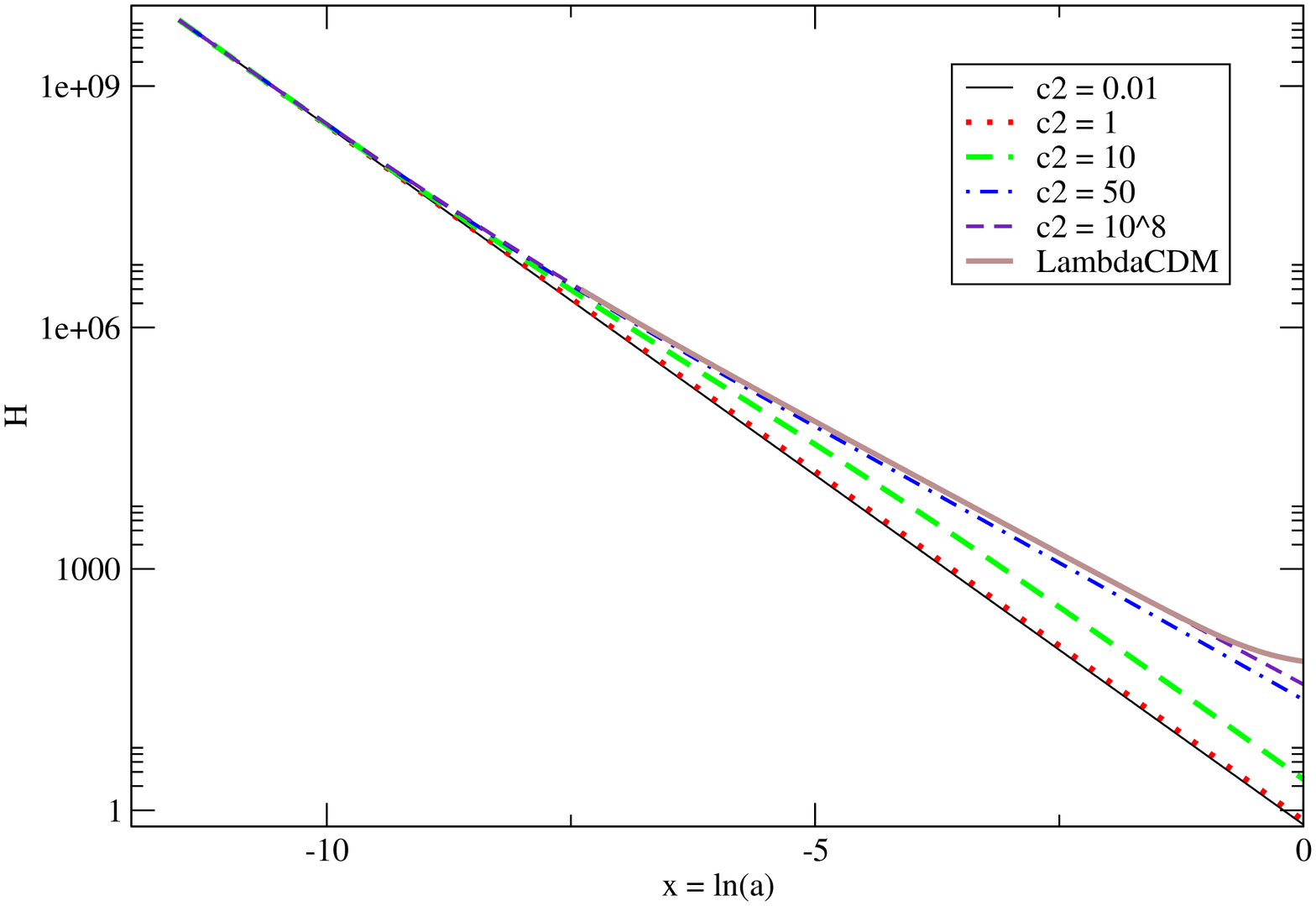}
      \caption{Plot of the evolution of the Hubble parameter $H$ in the galileon model with coefficients $c_1 = c_3 = 0$ and nozero $c_2$. The evolution of $H$ is shown for five different values of $c_2$: $c_2 = 0.01M_{\mathrm{PL}}^2$, $c_2 = 1M_{\mathrm{PL}}^2$, $c_2 = 10M_{\mathrm{PL}}^2$, $c_2 = 50M_{\mathrm{PL}}^2$ and $c_2 = 10^8M_{\mathrm{PL}}^2$. The evolution of $H$ for the usual $\Lambda$CDM model has also been plotted alongside the others.}
      \label{fig:BD_H}
  }

      \FIGURE{
      \includegraphics[width=0.6\textwidth]{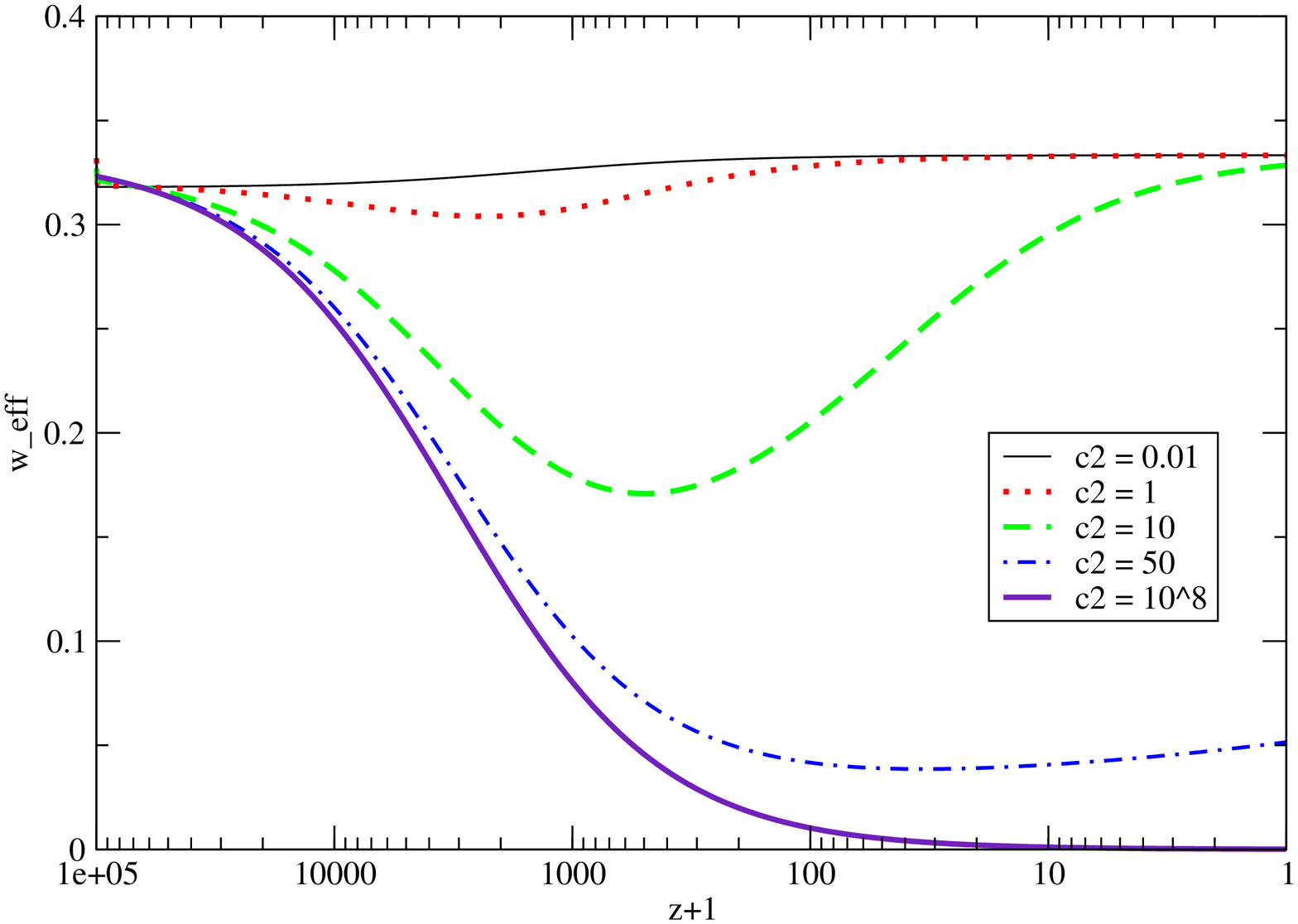}
      \caption{Plot of the evolution of the total equation of state parameter $w_{\mathrm{eff}}$ in the galileon model with coefficients  $c_1 = c_3 = 0$ and nozero $c_2$. The evolution is shown for five different values of $c_2$: $c_2 = 0.01M_{\mathrm{PL}}^2$, $c_2 = 1M_{\mathrm{PL}}^2$, $c_2 = 10M_{\mathrm{PL}}^2$, $c_2 = 50M_{\mathrm{PL}}^2$ and $c_2 = 10^8M_{\mathrm{PL}}^2$.}
      \label{fig:BD_w}
  }

  We see that $c_2$ in fact acts as a demarcation scale between the non-galileon background theory and the theory with a galileon with all the coefficients equal to zero. The larger the value of $c_2$, the later the galileon will have any impact on the evolution. However, as we raise the initial values of the galileon, the larger is the $c_2$ we need to compensate for this. Hence quite large initial values for the galileon will make the galileon dominate and give a Universe at early times with much smaller values of the Hubble parameter before transitioning into the radiation like phase. For instance in the case of $c_2 = 50M_{\mathrm{PL}}^2$ the initial values of $\pi$ and $\pi'$ were set to $0.5e-10$ and $0.5e-50$ respectively in the case shown in figures \ref{fig:BD_H} and \ref{fig:BD_w}. Changing the values to $0.5e-10$ and $0.5e-10$ respectively had very little effect, however a change to $0.5e-5$ and $0.5e-5$ for the initial values respectively changed the evolution of the background quite considerably, resulting in a much lower value for the Hubble constant, reaching a value comparable to that in the plotted case for $c_2 = 0.01M_{\mathrm{PL}}^2$ only very recently (redshift of only a small fraction), and from there transitioning into radiation like behaviour.

Since the theory only shows transitions between a pure radiation like model and the theory without galileons, this theory can not display selfacceleration. Hence for it to give accelerated expansion a cosmological constant or some other form of dark energy would need to be added. 

  That the impact of the galileon decreases for increasing values of $c_2$ can also be seen by comparing plots of the fractional densities and densities for two different values of $c_2$. This is shown for $c_2 = 10M_{\mathrm{PL}}^2$ and $10^8M_{\mathrm{PL}}^2$ in figures \ref{fig:BD_Om_10}, \ref{fig:BD_Om_108}, \ref{fig:BD_rho_10} and \ref{fig:BD_rho_108} respectively.

\FIGURE{
      \includegraphics[width=0.6\textwidth]{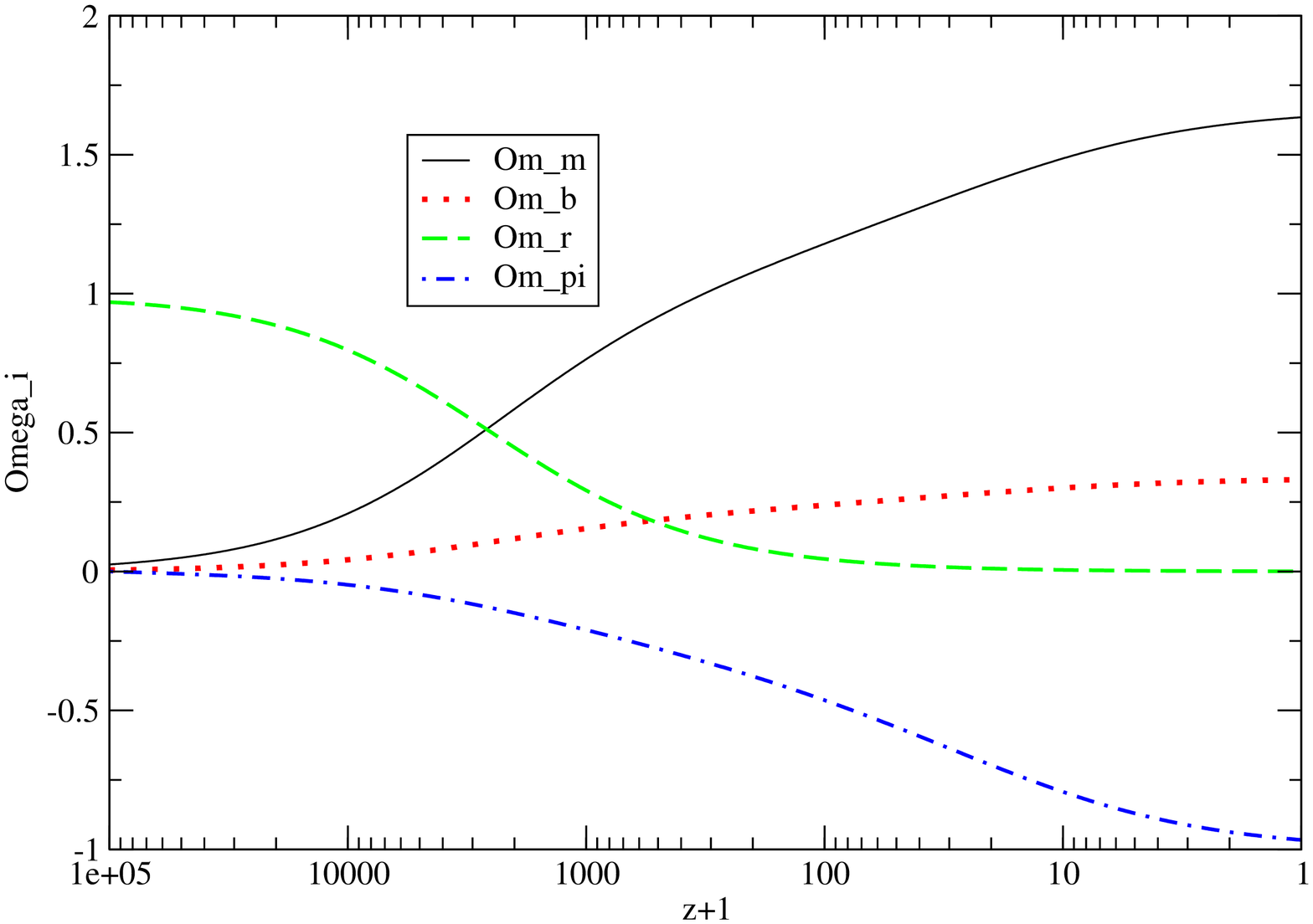}
      \caption{Plot of the evolution of the fractional densities $\Omega_i$ in the galileon model with coefficients $c_1 = c_3 = 0$, $c_2 = 10M_{\mathrm{PL}}^2$.}
      \label{fig:BD_Om_10}
  }

\FIGURE{
      \includegraphics[width=0.6\textwidth]{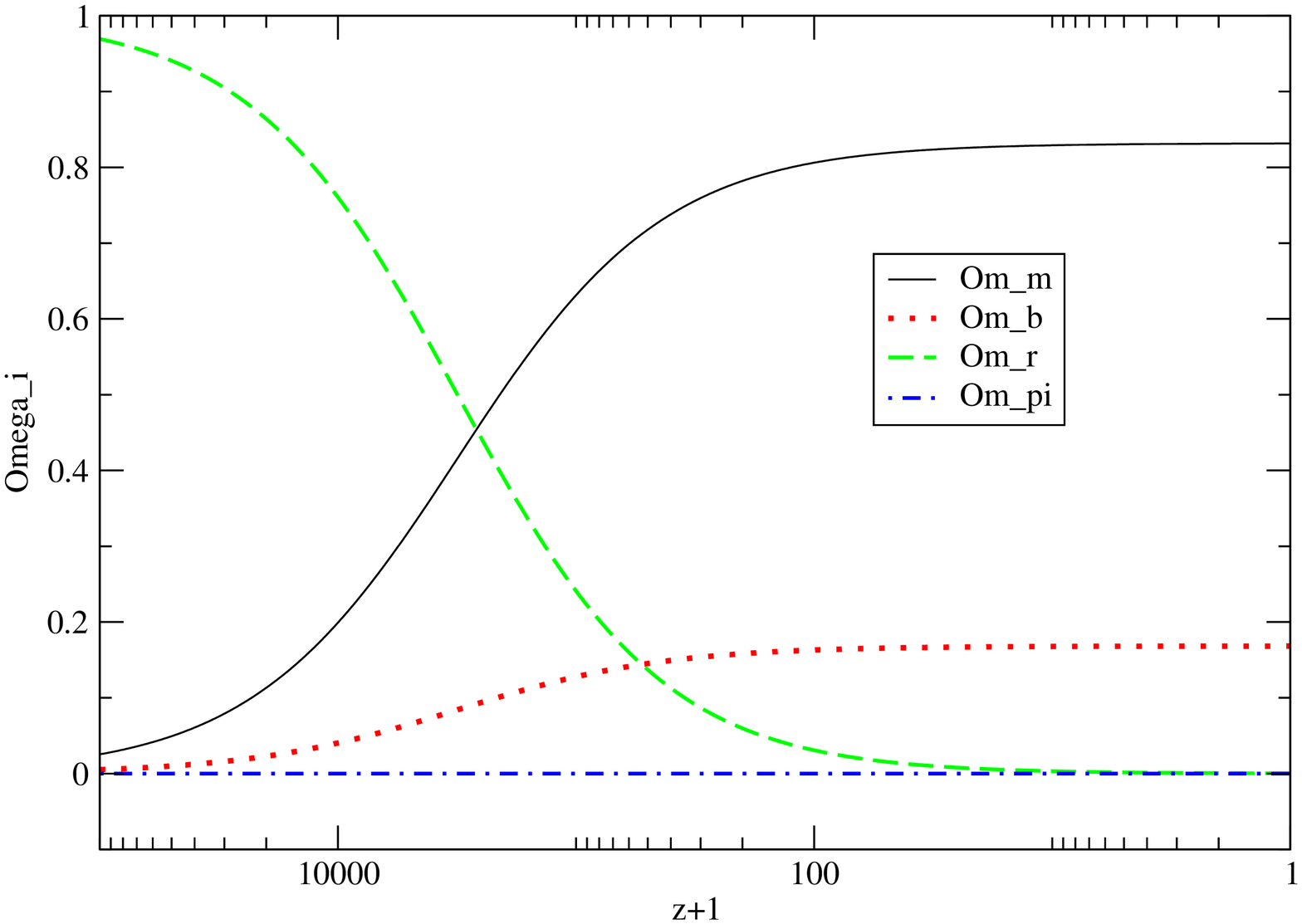}
      \caption{Plot of the evolution of the fractional densities $\Omega_i$ in the galileon model with coefficients $c_1 = c_3 = 0$, $c_2 = 10^8M_{\mathrm{PL}}^2$.}
      \label{fig:BD_Om_108}
  }

\FIGURE{
      \includegraphics[width=0.6\textwidth]{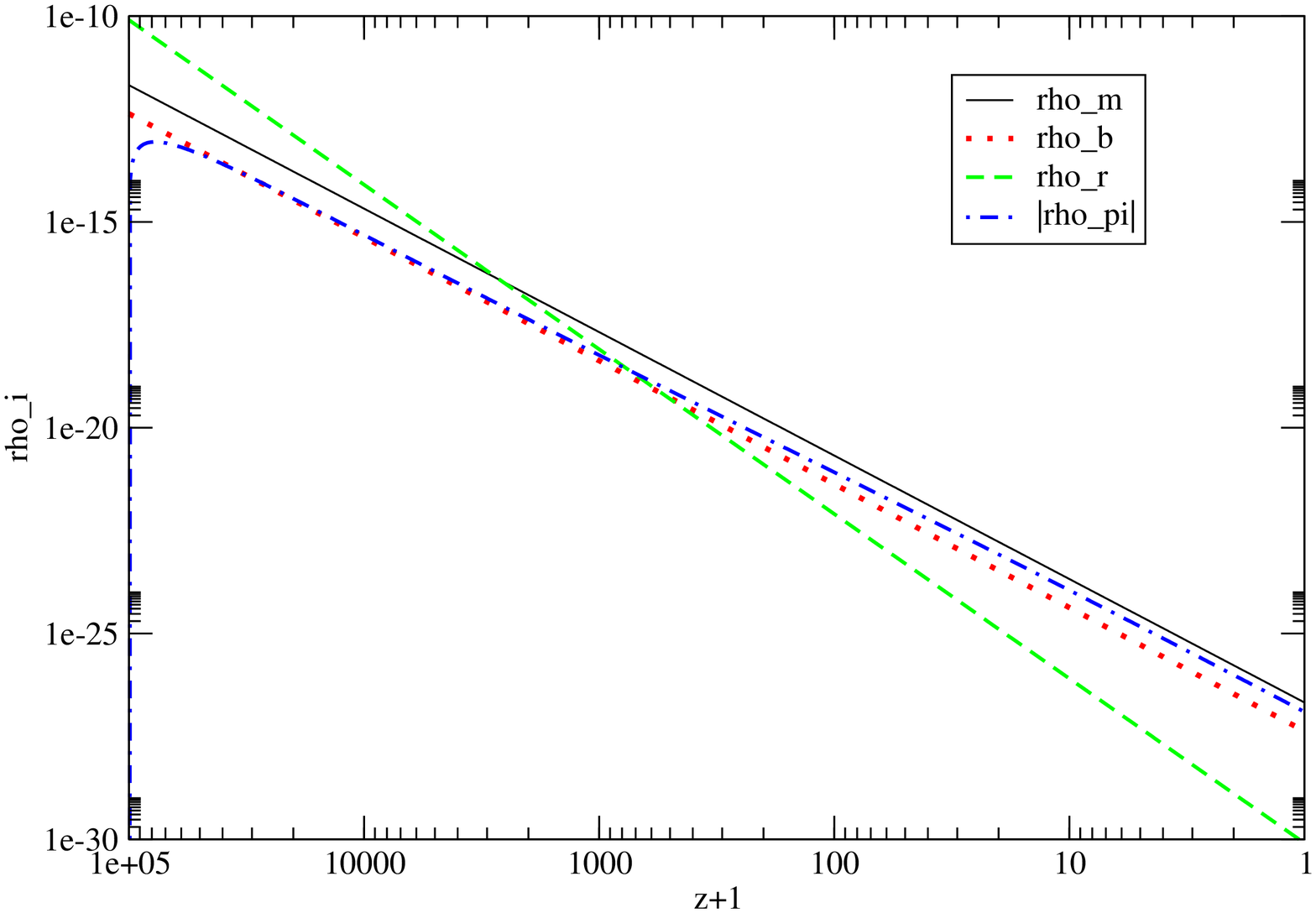}
      \caption{Plot of the evolution of the densities $\rho_i$ in the galileon model with coefficients $c_1 = c_3 = 0$, $c_2 = 10M_{\mathrm{PL}}^2$. Since the galileon density starts out negative its absolute value $|\rho_{\pi}|$ is shown in the plot.}
      \label{fig:BD_rho_10}
  }

\FIGURE{
      \includegraphics[width=0.6\textwidth]{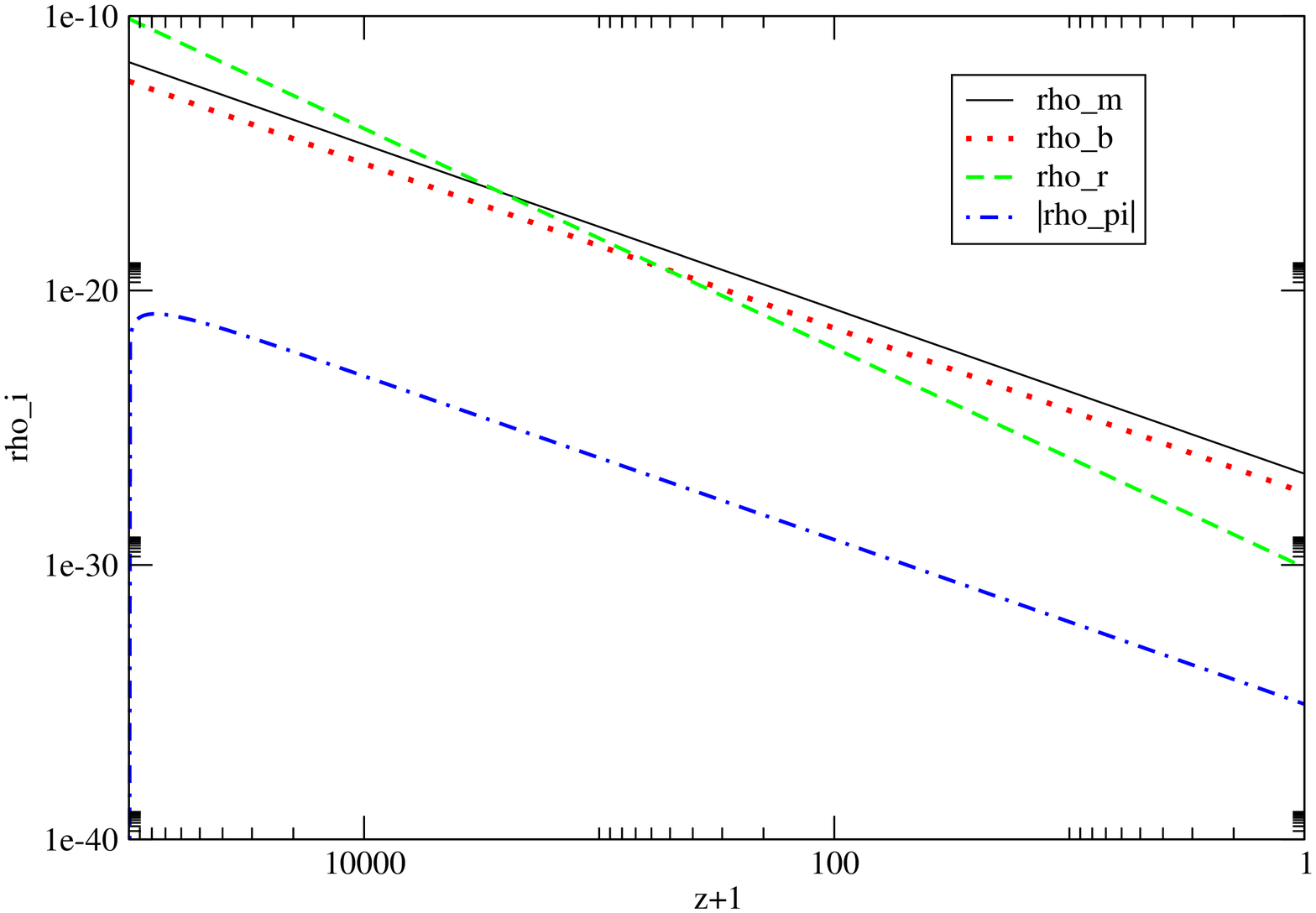}
      \caption{Plot of the evolution of the densities $\rho_i$ in the galileon model with coefficients $c_1 = c_3 = 0$, $c_2 = 10^8M_{\mathrm{PL}}^2$. Since the galileon density starts out negative its absolute value $|\rho_{\pi}|$ is shown in the plot.}
      \label{fig:BD_rho_108}
  }

  \subsection{Non-zero value of $c_3$ and added cosmological constant}\label{sec:ChowKhoury}
  The theory considered in \cite{Chow_et_Khoury} corresponds to a model of non-zero $c_3$, $c_1 = c_2 = 0$ with the addition of a cosmological constant. That is, a theory of a third order only galileon in a $\Lambda$CDM background.

  The equations in this case simplify significantly. The equations for the coefficients $\gamma$ and $\delta$ given in equation (\ref{eq:gamma_and_delta_general}) become:
   \begin{eqnarray}
     \gamma &=& \frac{M_{\mathrm{PL}}^2H^2e^{-2\pi}\left(\frac{3}{2} - 2\pi' + 2\left(\pi'\right)^2\right) + \frac{1}{2}P}{H^2\left(M_{\mathrm{PL}}^2e^{-2\pi} - \frac{c_3}{2}H^2\left(\pi'\right)^2\right)}\nonumber\\
     \delta &=&\frac{M_{\mathrm{PL}}^2e^{-2\pi}\left(1 - \pi'\right) + \frac{c_3}{2}H^2\left(\pi'\right)^3}{H\left(M_{\mathrm{PL}}^2e^{-2\pi} - \frac{c_3}{2}H^2\left(\pi'\right)^2\right)}
   \end{eqnarray}
   and the equation for $H'$ becomes:
   \begin{equation}
     H' = \frac{H^3c_3\pi'\left(2\gamma + 3\pi'\right) - 4HM_{\mathrm{PL}}^2e^{-2\pi}}{2M_{\mathrm{PL}}^2e^{-2\pi} - c_3H^2\pi'\left(3\pi' + 2H\delta\right)}
   \end{equation}

  We can obtain a model equal to the one considered in \cite{Chow_et_Khoury} by considering this regime with $c_1 = c_2 = 0$, $c_3 \neq 0$ in a background with a cosmological constant. However, since we had a slight discrepancy in the equations as compared to their results, there might be some differences. We must also pay attention to the units, as we are working with dimensionless $\pi$ fields, where as they have dimensionful $\pi$ fields.

  To review the results and compare to those of \cite{Chow_et_Khoury}, we have plotted the evolution of the Hubble parameter $H$ as compared to that of $\Lambda$CDM, the evolution of the energy densities $\rho_i$ and the evolution of the fractional densities $\Omega_i$ in figures \ref{fig:Chow_Khoury_like_H_100}, \ref{fig:Chow_Khoury_like_rhos_100} and in figure \ref{fig:Chow_Khoury_like_Omegas_correctL} for a particular value of $c_3$, $c_3 = 100M_{\mathrm{PL}}^2/H_0^2$. As in \cite{Chow_et_Khoury} we have defined
  \begin{equation}\label{eq:def_omegas}
    \Omega_i = \frac{\rho_ie^{2\pi}}{3H^2M_\mathrm{PL}^2}\mathrm{.}
  \end{equation}

  To make the results even more comparable to those in \cite{Chow_et_Khoury} we make the $\Omega_\Lambda$ value such that $\Omega_m$ today is close to the value measured today. We also extend the simulation to $z = -0.8$.

\FIGURE{
      \includegraphics[width=0.6\textwidth]{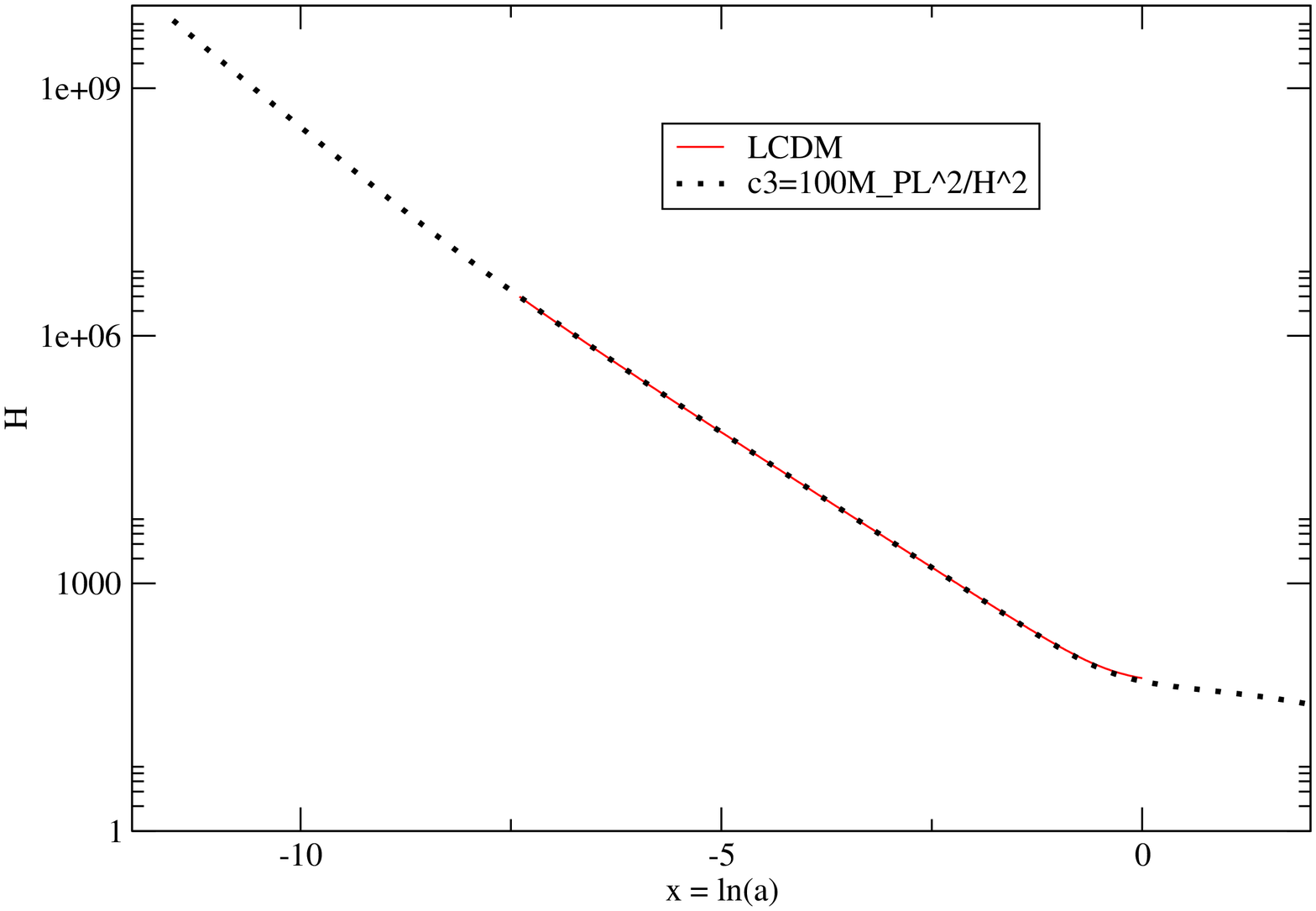}
      \caption{Plot of the evolution of the Hubble parameter $H$ in the galileon model with coefficients $c_1 = c_2 = 0$. $c_3 = 100M_{\mathrm{PL}}^2/H_0^2$. There is also a cosmological constant present in this model set so as to ensure $\Omega_m$ today of about the same size as measured. This corresponds quite closely to the case considered in \cite{Chow_et_Khoury}. The plot also shows the evolution of the Hubble parameter in the $\Lambda$CDM case.}
      \label{fig:Chow_Khoury_like_H_100}
  }

     \FIGURE{
      \includegraphics[width=0.6\textwidth]{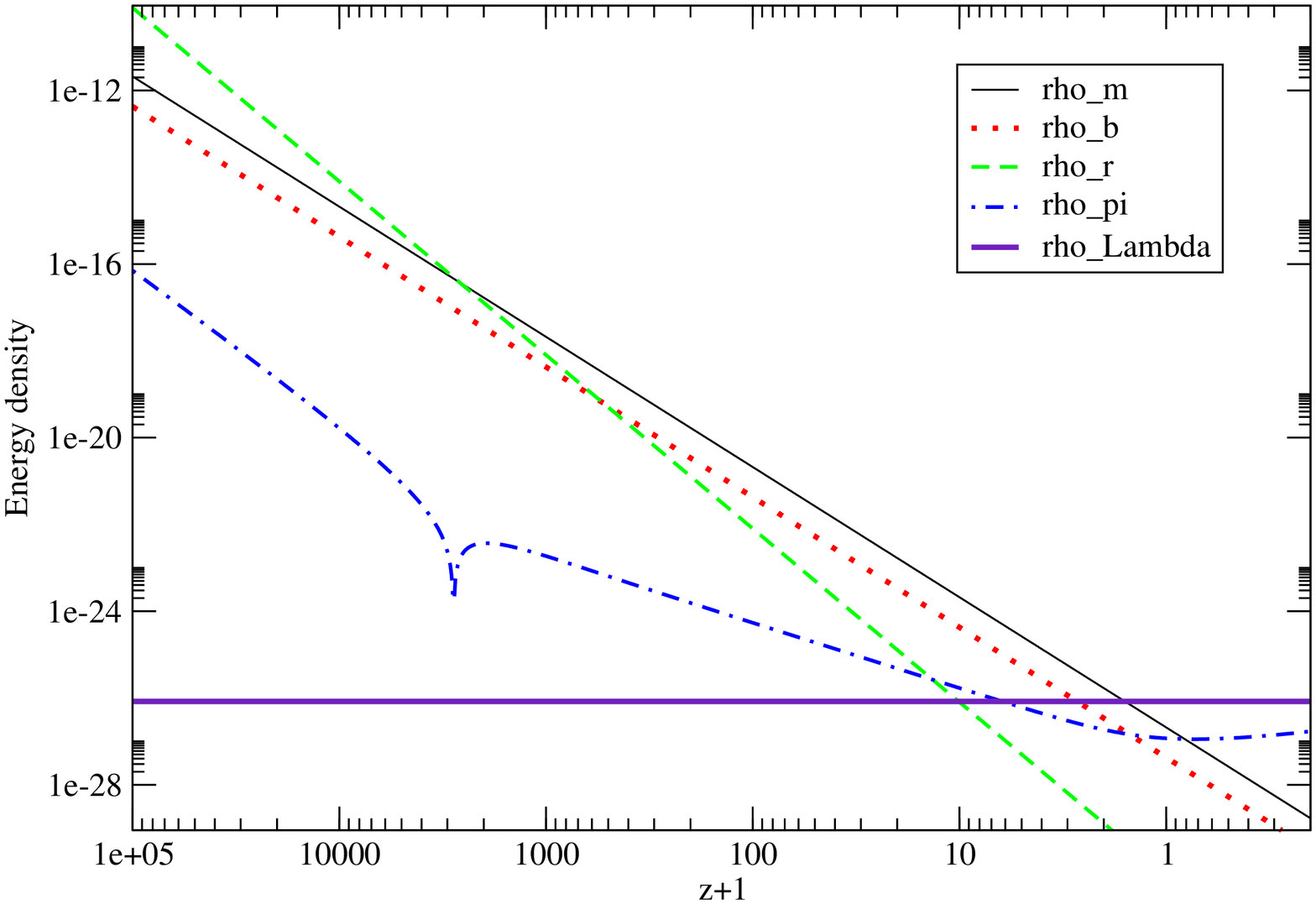}
      \caption{Plot of the evolution of the energy densities of the different components in the galileon model with coefficients $c_1 = c_2 = 0$, $c_3 = 100M_{\mathrm{PL}}^2/H_0^2$ and a cosmological constant.}
      \label{fig:Chow_Khoury_like_rhos_100}
  }

 \FIGURE{
      \includegraphics[width=0.6\textwidth]{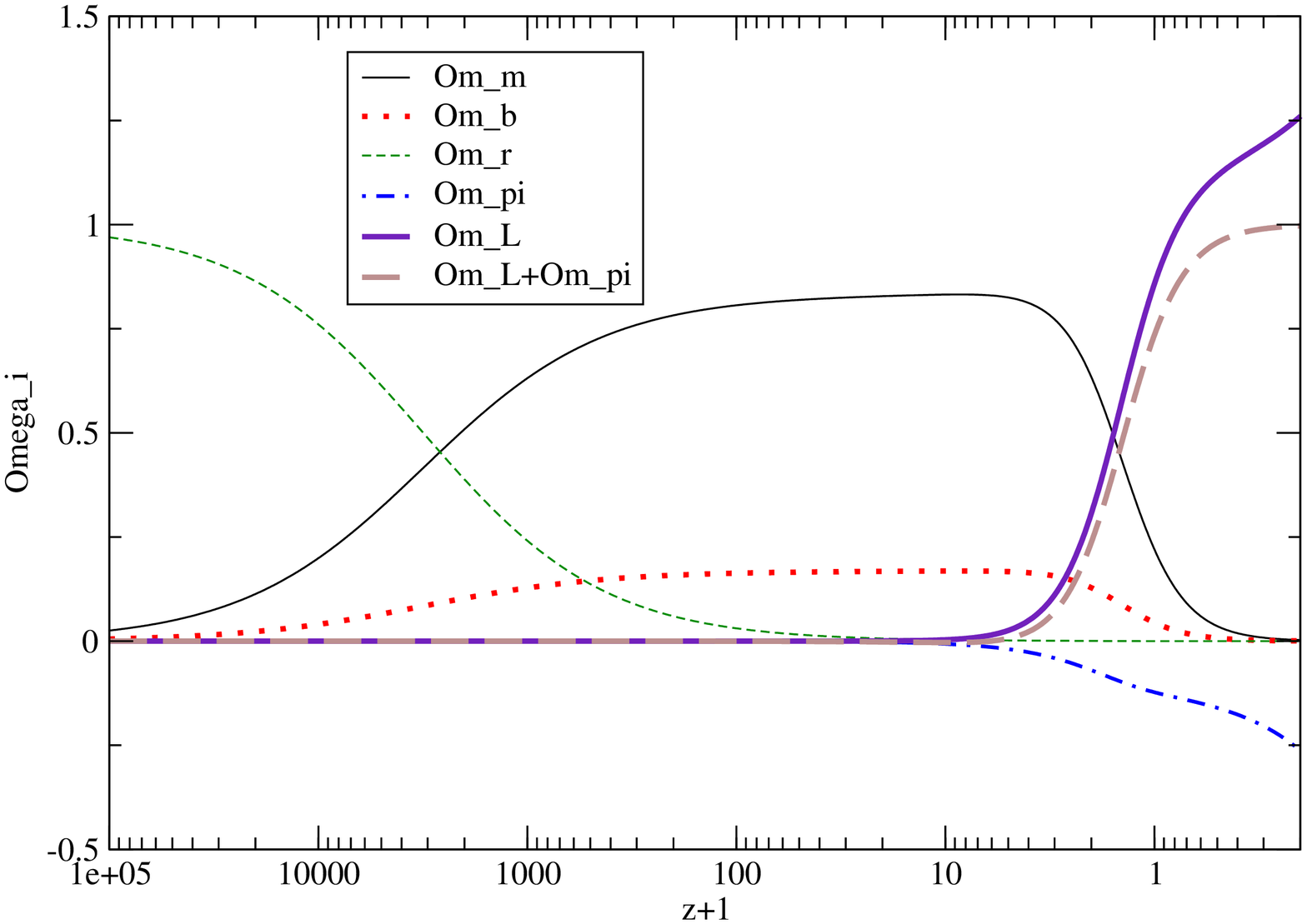}
      \caption{Plot of the evolution of the fractional energy densities of the different components in the galileon model with coefficients $c_1 = c_2 = 0$, $c_3 = 100M_{\mathrm{PL}}^2/H_0^2$. Here the added cosmological constant has been set so as to ensure $\Omega_m$ today close to its measured value, as in \cite{Chow_et_Khoury}.}
      \label{fig:Chow_Khoury_like_Omegas_correctL}
  }

  The results resemble closely the ones found in \cite{Chow_et_Khoury}. Similarly to \cite{Chow_et_Khoury} we also found that a correct evolution history was first obtained for relatively large values of $c_3$, the models resembling $\Lambda$CDM more the larger the value of $c_3$. Plots showing the evolution of the Hubble parameter and the evolution of the effective equation of state parameter in some different cases are given in figures \ref{fig:Chow_Khoury_likH_correctL} and \ref{fig:Chow_Khoury_like_w_correctL}. These plots show only small differences. 

In figures \ref{fig:Chow_Khoury_likH} and \ref{fig:Chow_Khoury_like_w} the difference in evolution histories is shown for even smaller values of $c_3$. In these plots the cosmological parameter has, however, only been set to its $\Lambda$CDM value, as normalising it in the very small $c_3$ cases is quite hard due to the large impact of the galileon field. 

The tendency is that for small enough values of $c_3$ the galileon field dominates the evolution too early for the Universe ever to experience a period of accelerated expansion. This should not come as a surprise after considering the case of all the coefficients set to zero in the previous subsection. In this case we have seen that the Universe will behave as if radiation dominated at all times no matter what the matter content of the Universe may be. The fact that the deviation from $\Lambda$CDM is larger with smaller $c_3$ is hence quite understandable, since $c_3$ is a demarcation scale, under which the physics still behaves as in a non-galileon Universe. The role of $c_3$ is hence quite similar to that of $c_2$ as described in section \ref{sec:Brans-Dicke} and we would expect the model with $c_2$ only with a cosmological constant added to have results somewhat comparable to the results found here.

\FIGURE{
      \includegraphics[width=0.6\textwidth]{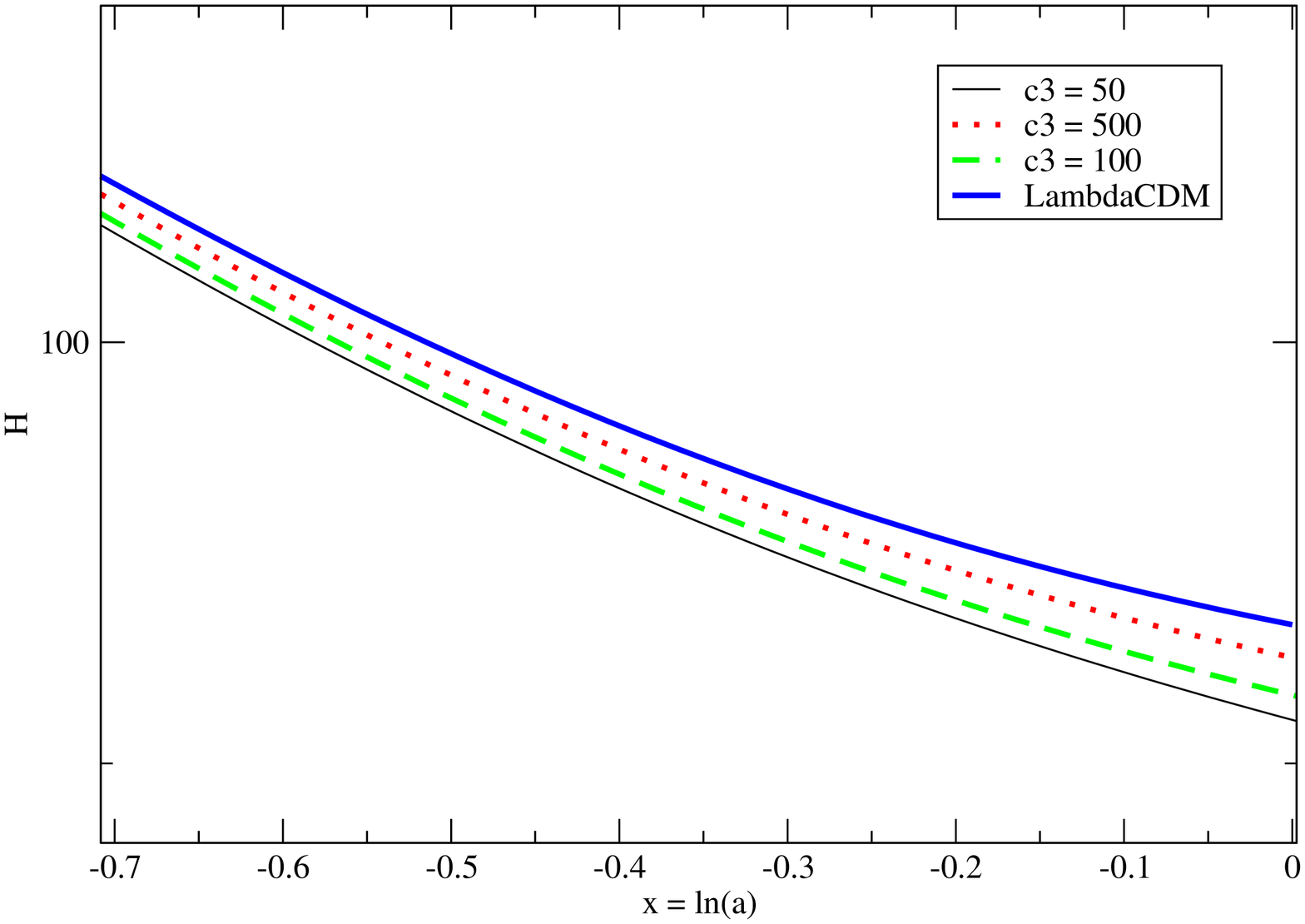}
      \caption{Plot of the evolution of the Hubble parameter $H$ in the galileon model with coefficients $c_1 = c_2 = 0$. $c_3$ is non zero and there is also a cosmological constant present in this model. Three cases are shown $c_3 = 50M_{\mathrm{PL}}^2/H_0^2$, $c_3 = 100M_{\mathrm{PL}}^2/H_0^2$ and $c_3 = 500M_{\mathrm{PL}}^2/H_0^2$. For comparison, the evolution of the Hubble parameter in the usual $\Lambda$CDM model is also shown in the plot.}
      \label{fig:Chow_Khoury_likH_correctL}
  }

     \FIGURE{
      \includegraphics[width=0.6\textwidth]{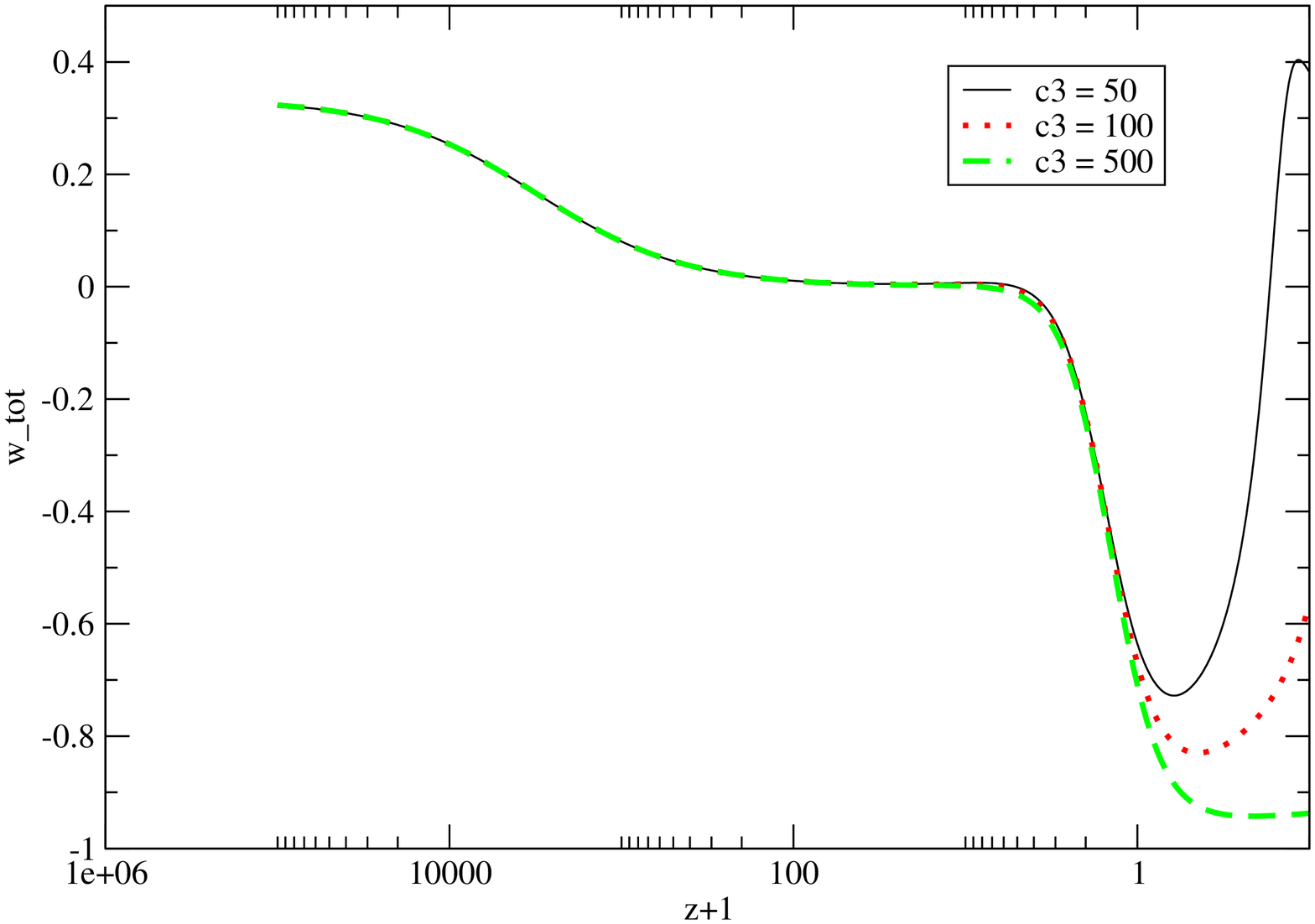}
      \caption{Plot of the evolution of the total effective equation of state parameter $w_{\mathrm{tot}}$ in the galileon model with coefficients $c_1 = c_2 = 0$. $c_3$ is non zero and there is also a cosmological constant present in this model. Three cases are shown $c_3 = 50M_{\mathrm{PL}}^2/H_0^2$, $c_3 = 100M_{\mathrm{PL}}^2/H_0^2$ and $c_3 = 500M_{\mathrm{PL}}^2/H_0^2$.}
      \label{fig:Chow_Khoury_like_w_correctL}
  }

  \FIGURE{
      \includegraphics[width=0.6\textwidth]{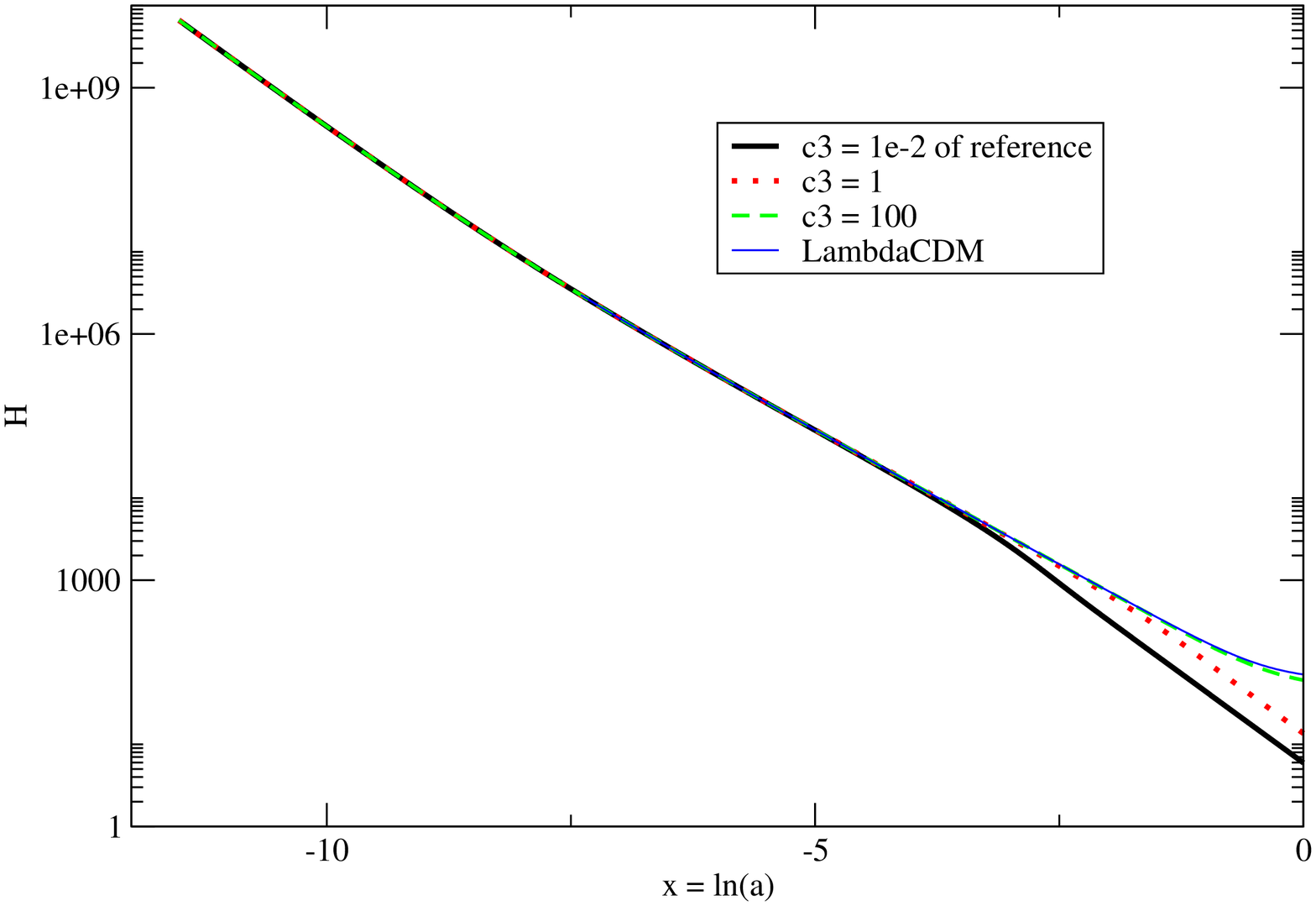}
      \caption{Plot of the evolution of the Hubble parameter $H$ in the galileon model with coefficients $c_1 = c_2 = 0$. $c_3$ is non zero and there is also a cosmological constant present in this model. This corresponds quite closely to the case considered in \cite{Chow_et_Khoury}, however, they have set the value of the cosmological constant so as to ensure a flat Universe today. Here I have simply used the cosmological constant value of the usual $\Lambda$CDM model. Three cases are shown: $c_3 = 0.01M_{\mathrm{PL}}^2/H_0^2$, $c_3 = M_{\mathrm{PL}}^2/H_0^2$ and $c_3 = 100M_{\mathrm{PL}}^2/H_0^2$. For comparison, the evolution of the Hubble parameter in the usual $\Lambda$CDM model is also shown in the plot.}
      \label{fig:Chow_Khoury_likH}
  }

     \FIGURE{
      \includegraphics[width=0.6\textwidth]{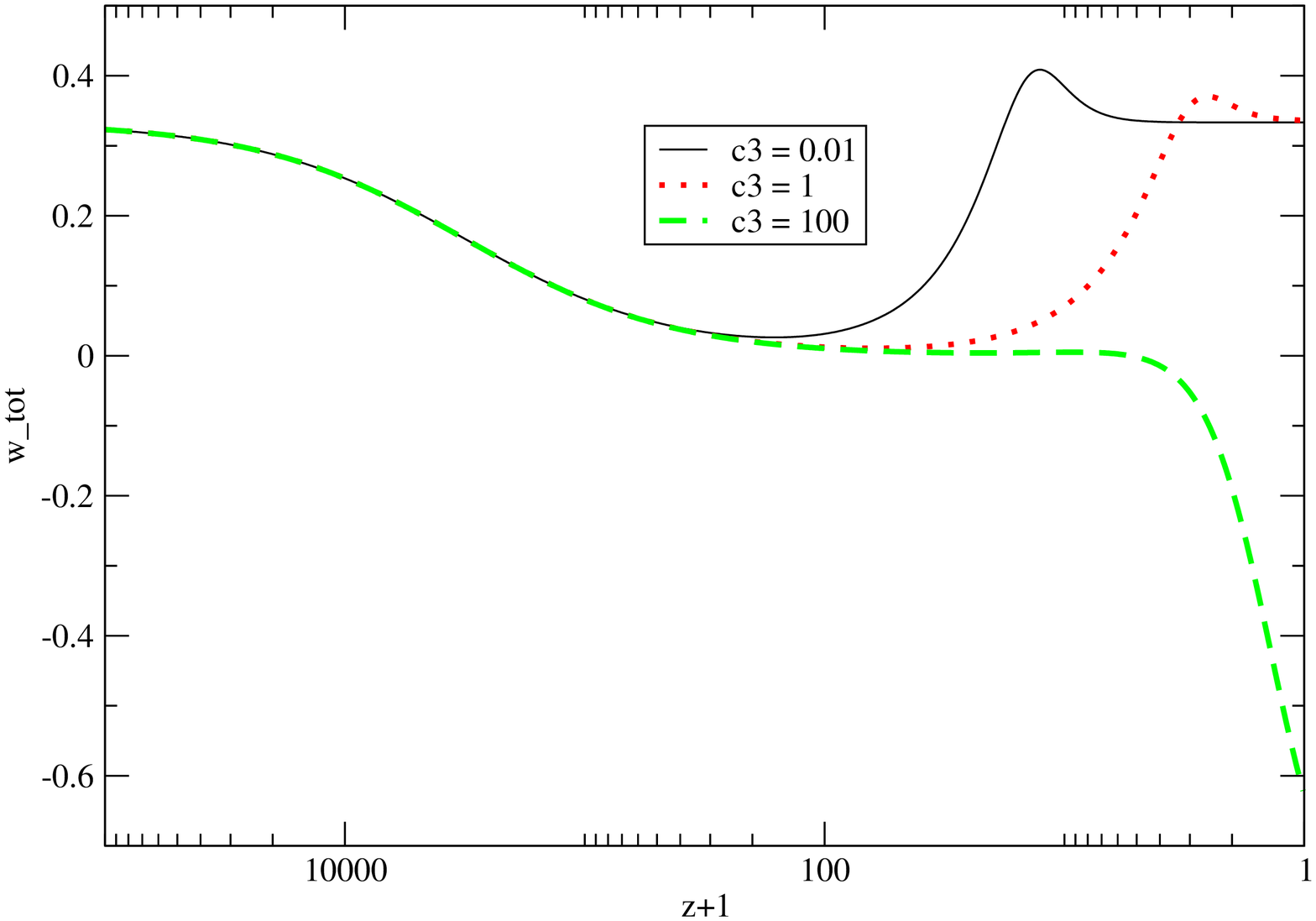}
      \caption{Plot of the evolution of the total effective equation of state parameter $w_{\mathrm{tot}}$ in the galileon model with coefficients $c_1 = c_2 = 0$. $c_3$ is non zero and there is also a cosmological constant present in this model. This corresponds quite closely to the case considered in \cite{Chow_et_Khoury}, however, they have set the value of the cosmological constant so as to ensure a flat Universe today. Here I have simply used the cosmological constant value of the usual $\Lambda$CDM model. Three cases are shown: $c_3 = 0.01M_{\mathrm{PL}}^2/H_0^2$, $c_3 = M_{\mathrm{PL}}^2/H_0^2$ and $c_3 = 100M_{\mathrm{PL}}^2/H_0^2$.}
      \label{fig:Chow_Khoury_like_w}
  }
  
  To check whether the behaviour is altered due to our slightly different results for the Raychaudhuri equation (\ref{eq:genRaychaudhuri}), we have performed simulations using their equation set as well as our own. They yield only slightly different results, i.e. the general behaviour is not changed and only very subtle changes in the evolution can be observed by reviewing the data very carefully.

  We have also considered the effect of altering the initial values of the galileon field. Actually, as long as the initial galileon field is small and positive $\pi_{\mathrm{start}} < 0.1$, and the initial galileon field prime derivative is quite small and negative $0 > \pi'_{\mathrm{start}} > -1e-8$, the evolution histories found for each value of $c_3$ is quite stable, that is with the initial values within these bounds, their exact values have little impact on the ensuing evolution history of the Universe. Relatively small initial values of the galileon field and its derivative seem like a reasonable starting point, so assuming this model to have the described behaviour seems fairly acceptable.

  \section{Numerical results for the full model background equations}

  Moving on to the full model, with $c_1$, $c_2$ and $c_3$ obeying the constraints of equation (\ref{eq:ConstraintsForFullModel}), which is supposed to show selfacceleration, we obtain an evolution for the Hubble parameter $H$ as shown in figure \ref{fig:full_H}. We see that the evolution of the Hubble parameter in this case is quite comparable to that of the $\Lambda$CDM model.

  \FIGURE{
      \includegraphics[width=0.6\textwidth]{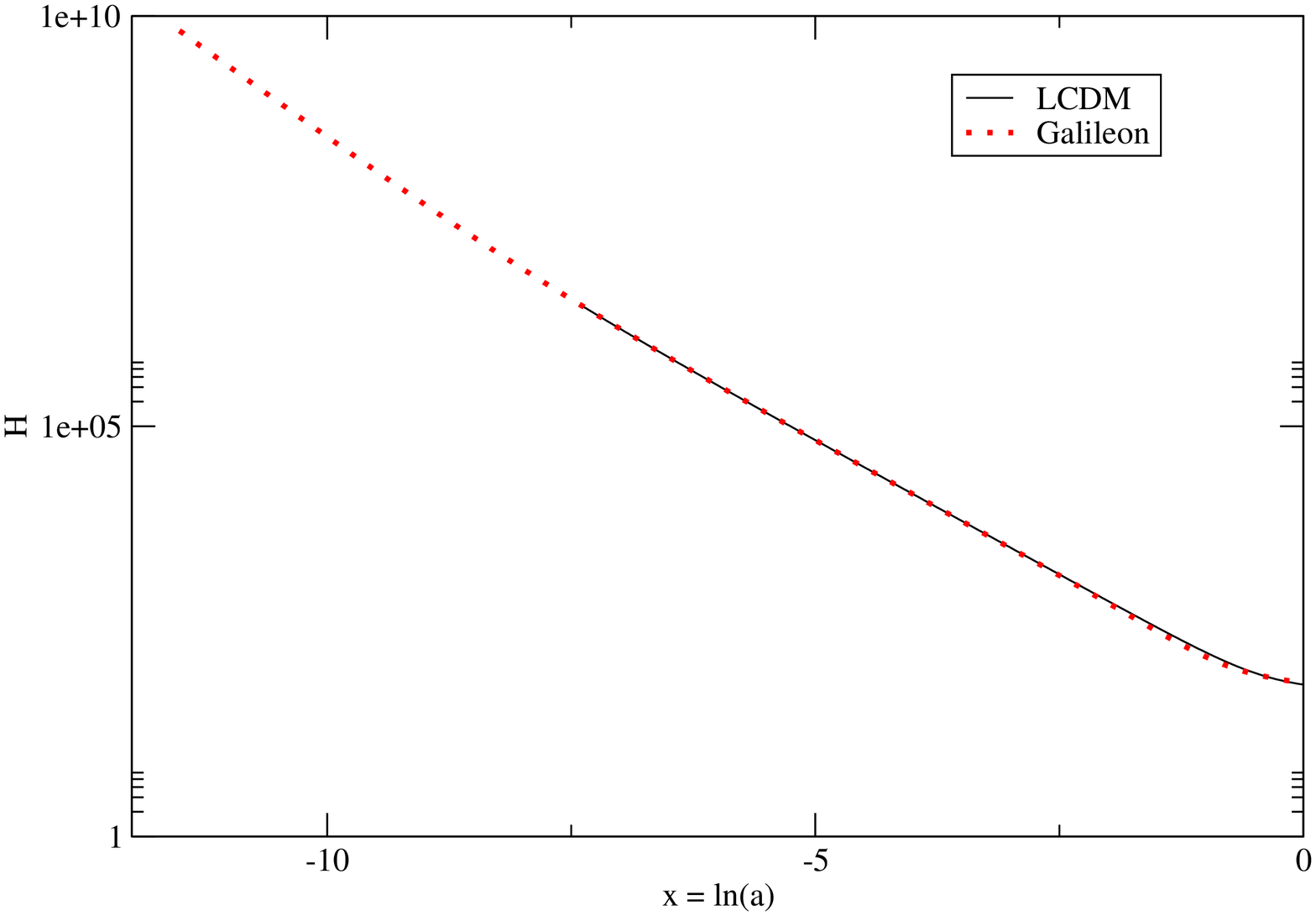}
      \caption{Plot of the evolution of the Hubble parameter $H$ in the case where $c_3 = 3M_{\mathrm{PL}}^2/H_0^2$, $c_2 = 18M_{\mathrm{PL}}^2$ compared to the evolution of the same parameter in the $\Lambda$CDM case. We see that the two evolution histories are quite comparable.}
      \label{fig:full_H}
}

  However, the evolution in this theory is not exactly as that of a $\Lambda$CDM model. This can be seen for instance from the evolution of the densities and fractional densities of the different components as shown in figures \ref{fig:full_rhos} and \ref{fig:full_Omegas}, or in the evolution of the equation of state parameter shown in figure \ref{fig:full_w_eff}.\footnote{The fractional densities are still defined as in equation (\ref{eq:def_omegas}).} This interesting behaviour may lead to possible ways of discerning this model from the $\Lambda$CDM model or at least to possibilities for constraining the values of $c_2$ and $c_3$ from distance measurements, such as supernova data or large scale structure surveys showing the evolution of the equation of state parameters over time.

  \FIGURE{
      \includegraphics[width=0.6\textwidth]{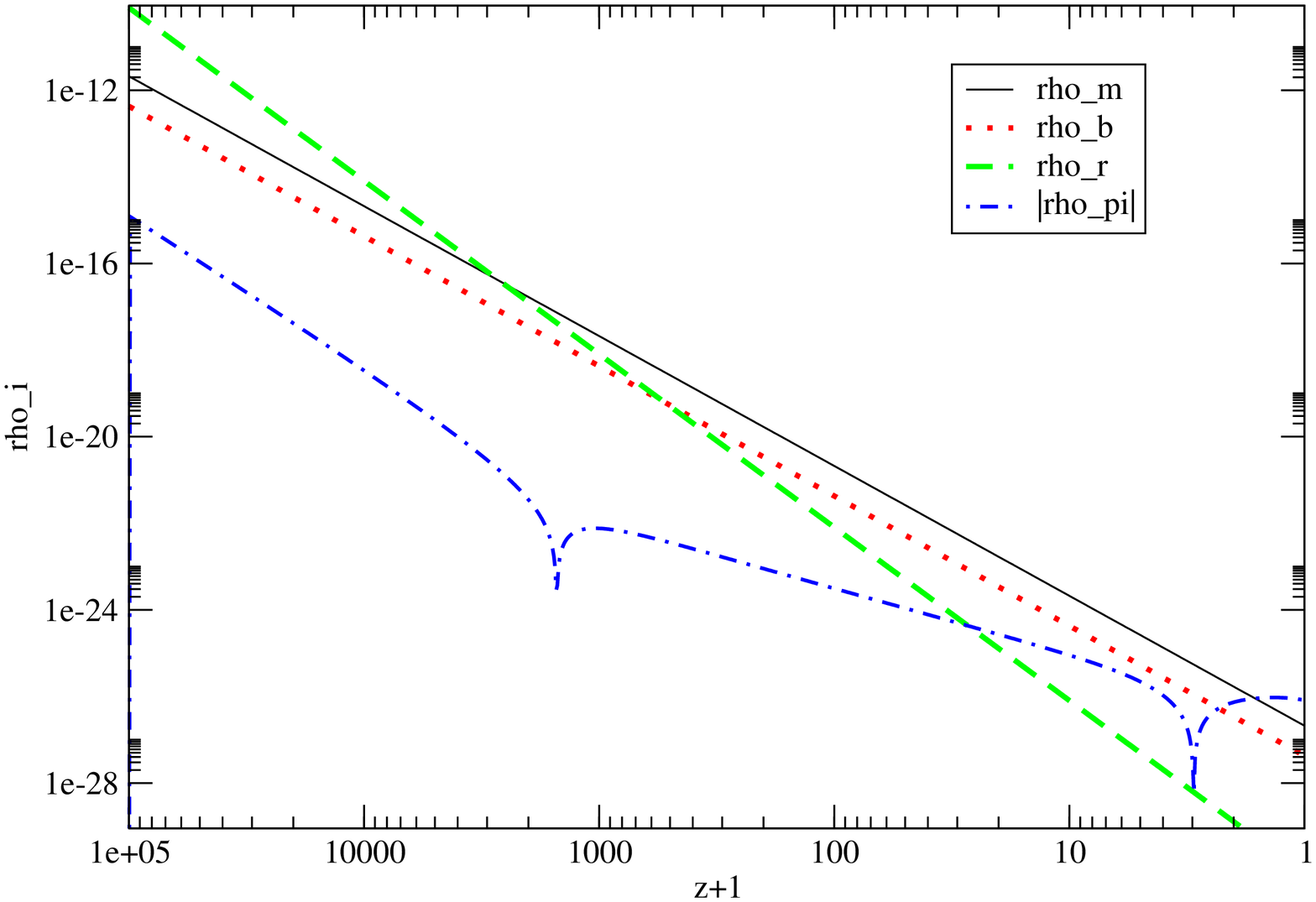}
      \caption{Plot of the evolution of the densities $\rho_i$ in the case where $c_3 = 3M_{\mathrm{PL}}^2/H_0^2$, $c_2 = 18M_{\mathrm{PL}}^2$.}
      \label{fig:full_rhos}
  } 

  \FIGURE{
      \includegraphics[width=0.6\textwidth]{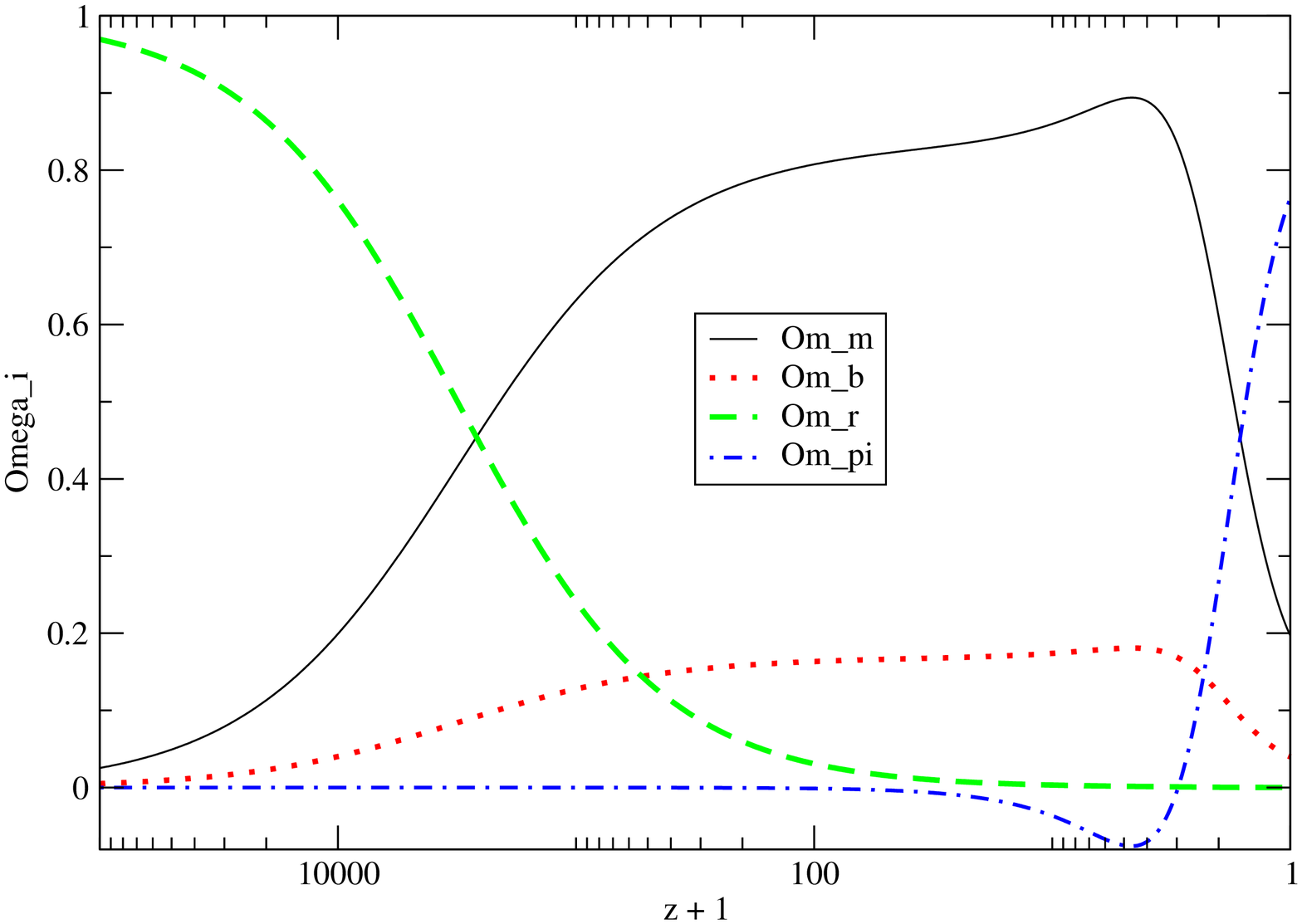}
      \caption{Plot of the evolution of the fractional densities $\Omega_i$ in the case where $c_3 = 3M_{\mathrm{PL}}^2/H_0^2$, $c_2 = 18M_{\mathrm{PL}}^2$.}
      \label{fig:full_Omegas}
  }

   \FIGURE{
      \includegraphics[width=0.6\textwidth]{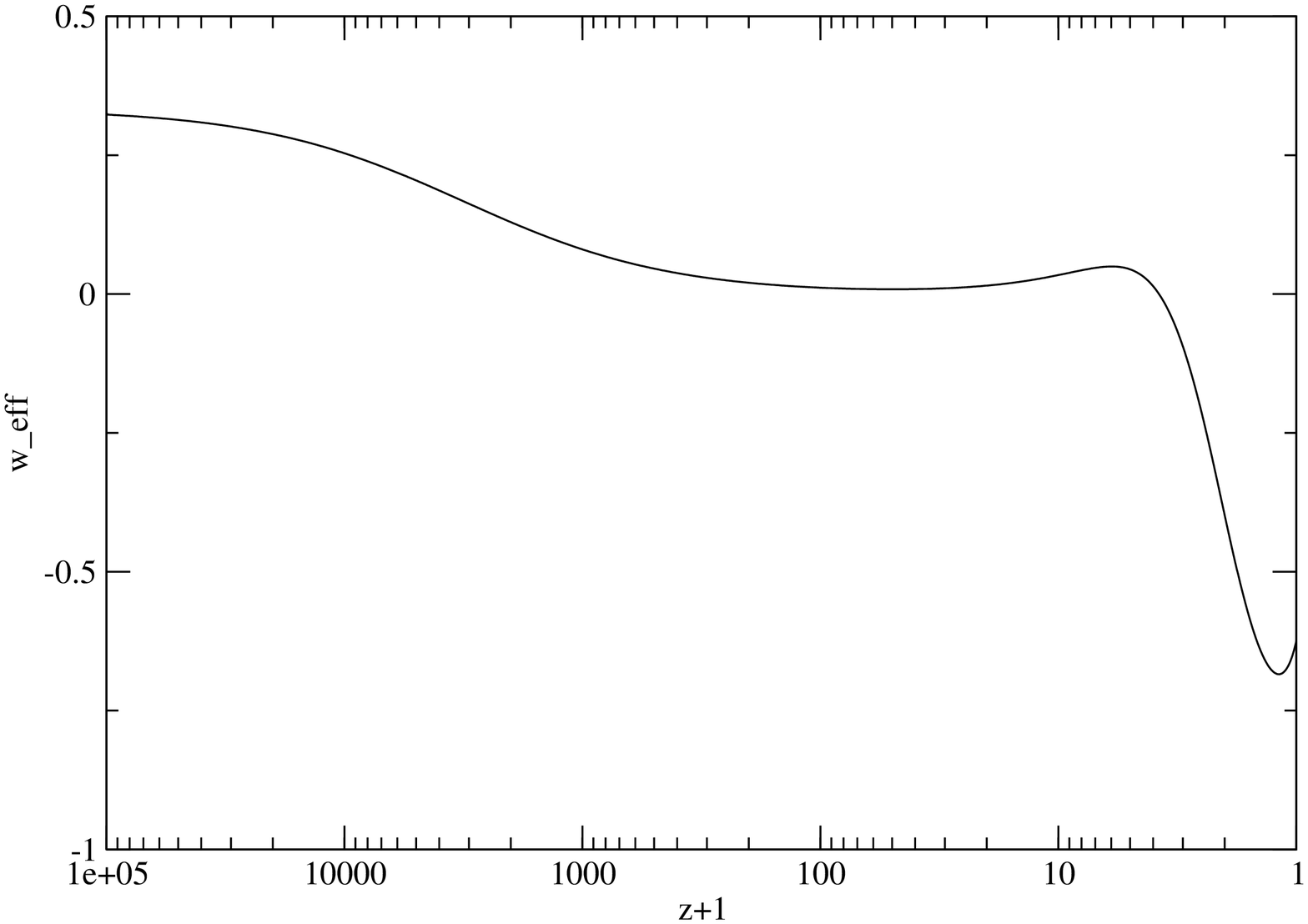}
      \caption{Plot of the evolution of the total equation of state parameter $w_{\mathrm{eff}}$ in the case where $c_3 = 3M_{\mathrm{PL}}^2/H_0^2$, $c_2 = 18M_{\mathrm{PL}}^2$. We see that after the radiation and matter dominated epochs, we get a slight increase in the effective equation of state parameter before it dips down to negative values. After this it starts increasing again. We can interpret the plot as follows. At early times the evolution follows that of the standard $\Lambda$CDM model, starting with radiation domination and then going into matter domination. As the galileon starts to dominate the evolution it has a negative density and is dominated by the coupling to gravity hence driving the theory towards a radiation dominated like epoch like in the all coefficients zero case. After only a short while of this however, the $c_1$ term takes over leading to a period of accelerated expansion. As time continues to go by we cross the demarcation scale given by the $c_2$ and $c_3$ entering an epoch where selfacceleration will stop and eventually turn to deceleration again.}
      \label{fig:full_w_eff}
  }

  \subsection{Robustness of results}
  
  Not surprisingly, the model does not stably yield near $\Lambda$CDM results under all initial values of the galileon field $\pi$ and its derivative. If these have values that are too large in the beginning, the early Universe evolution will be dominated by the galileon and hence not give results compatible with current knowledge from baryon nucleosynthesis etc.

  However, the evolution of the background equations also exhibit a dependence on the exact values of the coefficients $c_2$ and $c_3$. If $c_2$ is too small, and this seems to not only depend on the relative value with respect to $c_3$, the selfacceleration starts too late in the evolution of the Universe. If $c_2$ is too large the selfacceleration starts too early. The fine tuning needed for $c_2$ increases as $c_3$ gets much smaller or larger than a few $M_{\mathrm{PL}}^2/H_0^2$. We can summarise by saying that for the cosmological evolution to be close to our own, the values of the parameters need to be given by order a few times appropriate factors of $H_0$ and $M_{\mathrm{PL}}$. This makes it tempting to say that although the parameter space is constrained, the allowed values constitute a natural parameter regime.

  After considering the galileon model with each term separately in sections \ref{sec:Tadpole} to \ref{sec:ChowKhoury} this is quite understandable. Since $c_2$ and $c_3$ are demarcation scales, keeping the theory away from the pure radiation dominated like Universe, at least one of these must be not too small to yield sensible results. Since our stability conditions dictate that $c_2 > 3H_0^2c_3$, we get that $c_2$ must be large enough. Since $c_1$ is the only term that can in fact yield selfacceleration, this term must be present, and to get the right value of the Hubble parameter today, and the right evolution, $c_1$ must be set by the demarcation scales. Having the demarcation scales both too large also causes trouble, since $c_1$ then gets very large and starts the accelerated expansion too early.

  For the initial values of the galileon and its derivative we find that they must be tuned to be quite close to zero. In the case of $c_3 = 3M_{\mathrm{PL}}^2/H_0^2$, $c_2 = 18M_{\mathrm{PL}}^2$, the evolution remains stable as long as the initial value for $\pi' < 1e-20$ with the initial value of $\pi < 1e-9$. As we change only the initial value of $\pi$ to around $1e-8$, the evolution shortens its matter dominated epoch and gets a more abrupt change to the selfaccelerated phase. Outside this regime the evolution becomes dominated by the galileon from a very early stage, and it becomes totally unstable. This is similar to the results found for perturbations in \cite{Kobayashi_et_al2010a}, where the initial values for the galileon had to be tuned to zero in the beginning in order for the galileon fluctuations not to take over and spoil the homogeneity and isotropy of the background.

  \section{Discussion}
  In this paper we have explored a certain version of the galileon model formulated in \cite{Nicolis_et_al2008}. Using their derivation of stability and selfaccelerated expansion, we have found a model that both admits selfacceleration and gives stable spherically symmetric solutions, that is a third order galileon with certain constraints on the three parameters. This model has not been thoroughly studied in the literature, and we feel that it is important to provide such a study of one of the simplest proposals for a selfaccelerating galileon.

  We have explored the background cosmology of this model, using numerical simulations of each term separately to build intuition. In this way we found that it is the tadpole term that leads to the selfacceleration of the solution. The second and third order terms defines a sort of demarcation scale above which the selfacceleration slows down, creating an evolution history not totally equal to that of $\Lambda$CDM. We also found that the galileon density becomes negative before it starts dominating. As it passes through zero and on to positive values the evolution of the equation of state goes through a phase where it grows slightly. This is a feature of the chosen coupling to gravity, which we have seen yields radiation dominated like solutions if present with no other galileon dynamics.

  We also found that the parameter space for a solution comparable to $\Lambda$CDM is limited, but not totally constrained. The bounds on the parameters to give a $\Lambda$CDM like evolution history are, however, consistent with solar system and galaxy cluster gravitational bounds found in \cite{Nicolis_et_al2008} and they lie in a somewhat natural range. We presume that further studies on the perturbations of this model will set even more stringent bounds on these parameters.

  We also found that the initial values of the galileon field and its derivative must be tuned to values quite close to zero. With larger values of these initial field, the model becomes highly unstable as the galileon begins to dominate quite early, leading to cosmological histories very different from the one favoured by present day observations. This is similar to results found in \cite{Kobayashi_et_al2010a} where the galileon perturbations had to be tuned to exactly zero in the beginning to avoid a total destruction of the homogeneity and isotropy of the background space. Further studies on the perturbations of this model will be interesting in finding out whether the same kind of anisotropic stress will be found here.

  The growth of structure in the galileon Universe will also be of interest as it might provide nice signatures in both baryonic acoustic oscillations and CMB, and these can only be found through studies of the perturbation equations of this model, which we postpone for a future publication.

  \begin{acknowledgments}
    We would like to thank Justin Khoury for useful discussions. DFM thanks the Research Council of Norway for the FRINAT grant 197251/V30.
  \end{acknowledgments}

  \bibliographystyle{JHEP}
  \bibliography{sources}

\end{document}